\documentclass[twocolumn]{aastex631}

\newcommand{\clustername}{SPT-0356}
\newcommand{\clusterfullname}{SPT-CLJ\,0356$-$5337}
\newcommand{\clusterz}{1.034} 
\newcommand{\kms}{km s$^{-1}$}
\newcommand{\z}{\textit{z}}
\newcommand{\msun}{${\rm M}_\odot$}

\definecolor{LightCyan}{rgb}{0.88,1,1}

\newcommand{\done}{\textcolor[rgb]{0,0,0}}
\newcommand{\revision}[1]{#1}
\newcommand{\rerevision}[1]{#1}

\newcommand{\Lya}{Ly$\alpha$}
\newcommand{\lya}{Ly$\alpha$}

\newcommand{\zsystemIII}{3.048}
\newcommand{\zsystemIV}{3.0205}
\newcommand{\zsystemVII}{5.3288}
\newcommand{\zsystemX}{4.1448}
\newcommand{\muLAEnorth}{{$\mu_{med}=3.9\pm 0.6$}} 
\newcommand{\muLAEmiddle}{{$\mu_{med}=3.7\pm 0.5$}} 
\newcommand{\muLAEsouth}{{$\mu_{med}=2.0\pm 0.3$}} 

\newcommand{\lenstool}{{\tt{Lenstool}}}

\newcommand{\Lenstool}{{\tt{Lenstool}}}

\newcommand{\massclusterB}{M($<$500 kpc)=$3.75^{+0.4}_{-0.34} \times 10^{14}$M$_{\odot}$}   
\newcommand{\massBCGB}{M$_{BCG}(<80 \text{kpc})= 3.93^{+0.21}_{-0.14}    \times 10^{13} $M$_{\odot}$}
\newcommand{\massLRGsB}{M$_{LRG}(<80 \text{kpc})= 2.92^{+0.16}_{-0.23}    \times 10^{13} $M$_{\odot}$}

\newcommand{\massBCGMB}{M$_{BCG}(<80 \text{kpc})= 4.04^{+0.07}_{-0.10}    \times 10^{13} $M$_{\odot}$}
\newcommand{\massLRGsMB}{M$_{LRG}(<80 \text{kpc})= 3.41\pm0.09    \times 10^{13} $M$_{\odot}$}

\newcommand{\massLRGsMD}{M$_{LRG}(<80 \text{kpc})= 3.01^{+0.42}_{-0.25}    \times 10^{13} $M$_{\odot}$}

\begin{document}

\title{The $z\sim1.03$ Merging Cluster SPT-CL J0356–5337: New  Strong Lensing Analysis with HST and MUSE }

\author[0000-0003-3971-5727]{Grace Smith}
\affiliation{Department of Astronomy, University of Michigan, 1085 S. University Ave, Ann Arbor, MI 48109, USA}

\author[0000-0003-3266-2001]{Guillaume Mahler}
\affiliation{STAR Institute, Quartier Agora - All\'ee du six Ao\^ut, 19c B-4000 Li\`ege, Belgium}
\affiliation{Centre for Extragalactic Astronomy, Durham University, South Road, Durham DH1 3LE, UK}
\affiliation{Institute for Computational Cosmology, Durham University, South Road, Durham DH1 3LE, UK}

\author[0000-0003-4470-1696]{Kate Napier}
\affiliation{Department of Astronomy, University of Michigan, 1085 S. University Ave, Ann Arbor, MI 48109, USA}

\author[0000-0002-7559-0864]{Keren Sharon}
\affil{Department of Astronomy, University of Michigan, 1085 S. University Ave, Ann Arbor, MI 48109, USA}

\author[0000-0003-1074-4807]{Matthew Bayliss} 
\affiliation{Department of Physics, University of Cincinnati, Cincinnati, OH 45221, USA}

\author[0000-0002-5108-6823]{Bradford Benson}
\affiliation{Fermi National Accelerator Laboratory, MS209, P.O. Box 500, Batavia, IL 60510, USA}
\affiliation{Department of Astronomy and Astrophysics, University of Chicago, 5640 South Ellis Avenue, Chicago, IL 60637, USA}

\author[0000-0001-7665-5079]{Lindsey Bleem}
\affiliation{High-Energy Physics Division, Argonne National Laboratory, 9700 South Cass Avenue., Lemont, IL, 60439, USA}
\affiliation{Kavli Institute for Cosmological Physics, University of Chicago, 5640 South Ellis Avenue, Chicago, IL, 60637, USA}

\author[0000-0003-4175-571X]{Benjamin Floyd} 
\affiliation{Faculty of Physics and Astronomy, University of Missouri-Kansas City, 5110 Rockhill Road, Kansas City, MO 64110, USA}
\affiliation{Institute of Cosmology and Gravitation, University of Portsmouth, Dennis Sciama Building, Portsmouth PO1 3FX, UK}

\author[0000-0003-1370-5010]{Michael D. Gladders} 
\affiliation{Department of Astronomy and Astrophysics, University of Chicago, 5640 South Ellis Avenue, Chicago, IL  60637, USA}
\affiliation{Kavli Institute for Cosmological Physics, University of Chicago, Chicago, IL 60637, USA}

\author[0000-0002-3475-7648]{Gourav Khullar} 
\affiliation{Department of Physics and Astronomy, and PITT PACC, University of Pittsburgh, Pittsburgh, PA 15260, USA}

\author[0000-0002-6987-7834]{Tim Schrabback} 
\affiliation{Universitat Innsbruck, Institut fur Astro- und Teilchenphysik, Technikerstr. 25/8, 6020 Innsbruck, Austria}

\begin{abstract} 
We present a strong lensing analysis and reconstruct the mass distribution of SPT-CL\,J0356$-$5337, a galaxy cluster at redshift $z=\clusterz$.
Our model supersedes previous models by making use of new multiband HST data and MUSE spectroscopy. We identify two additional lensed galaxies to inform a more well-constrained model using 12 sets of multiple images in five separate lensed sources. The three previously known sources were spectroscopically confirmed by \cite{Mahler2020} at redshifts of \z\ = 2.363, \z\ = 2.364, and \z\ = 3.048. We measured the spectroscopic redshifts of two of the newly discovered arcs using MUSE data, at \z\ = 3.0205 and \z\ = 5.3288. We increase the number of cluster member galaxies by a factor of three compared to previous work. We also report the detection of extended \lya\ emission from several background galaxies. 
{We measure the total projected mass density of the two major sub-cluster components, one dominated by the BCG, and the other by a compact group of luminous red galaxies. We find \massBCGB\ and \massLRGsB, yielding a mass ratio of $1.35 ^{+0.16}_{-0.08}$}. 
The strong lensing constraints offer a robust estimate of the projected mass density regardless of modeling assumptions; allowing more substructure in this line of sight does not change the results or conclusions. 
Our results corroborate the conclusion that SPT-CL\,J0356$-$5337 is dominated by two mass components and is likely undergoing a major merger on the plane of the sky. 
\end{abstract}


\section{Introduction}
\label{sec:intro}
\done{Galaxy clusters, especially those at high redshift, probe large-scale structure formation. Clusters grow by undergoing mergers with other mass components. Whereas minor mergers, involving galaxy-to-group-scale halos falling into the cluster-scale halo, are rather common, major mergers---where the merging component is at least one third of the mass of the cluster---are much rarer \citep{Fakhouri2008}. Mergers of clusters represent short events in the long history of the clusters, which offer insight into the buildup of large-scale structure in the Universe, and the growth and redistribution of dark matter and baryons in the densest nodes of the cosmic web. 
Studying this phase and separating the masses of the two (or more) merging components requires a high resolution mass probe capable of measuring of the mass distribution \revision{within radii smaller than 100-500 kpc }
of the cluster core. The best probe at such scales is strong gravitational lensing, which can map the projected mass density with high resolution.}

\done{There are only a few known major mergers that exhibit strong-lensing evidence, which can be used to map in detail the dark matter at the core of the merger. Some well-studied examples include the ``Bullet Cluster" (1E 0657$-$558, \citealt{Bradac2006, Clowe2006}) and ``El Gordo" (ACT$-$CL J0102-4915, \citealt{Menanteau2012, Caminha2023}). While rare, the multiwavelength studies of such lines of sight advanced our understanding of some of the fundamental properties of the Universe, such as providing solid evidence for dark matter \citep[e.g.,][]{Clowe2006} and constraining the dark matter self-interaction cross section \citep[e.g.,][]{Harvey2015,Markevitch2004,Wittman2018,Randall2008,Bradac2008}.}

\done{The subject of this paper is \clusterfullname\ (hereafter \clustername), a galaxy cluster at redshift $z=\clusterz$ acting as a strong gravitational lens.
The cluster was discovered by the South Pole Telescope (SPT) based on its Sunyaev-Zel'dovich (SZ) effect signature \citep{Bleem2015,Bleem2020}. It is reported to have a total mass of ${\rm M}_{200c}=5.12^{+1.00}_{-1.32} \times 10^{14}$ \msun $h_{70}^{-1}$ and ${\rm M}_{500c} = 3.59^{+0.59}_{-0.66} \times 10^{14}$ \msun $h_{70}^{-1}$ \citep{Bocquet2019}.
Optical imaging and spectroscopic follow-up of \clustername\ \citep{Mahler2020} revealed that it is composed of two main mass components at the same redshift, separated by $\sim$170 kpc: the eastern component is associated with the brightest cluster galaxy (BCG), and the western component is associated with a group of luminous red galaxies (LRGs). 
The two-component mass distribution, paired with the fact that these two components have nearly equal radial velocities (within 300 km s$^{-1}$), implies that \clustername\ is undergoing a merger on the plane of the sky.
Analysis of the mass distribution would determine whether the system could be classified as a major merger. } 

\done{In the first lensing analysis of \clustername, \cite{Mahler2020} reported that the mass distribution in the lens plane is consistent with two major cluster-scale halos, with a mass ratio between 1:1.25 and 1:1.58. Their lens model relied on ground-based imaging and spectroscopy and shallow single-band Hubble Space Telescope (HST) imaging data, and was underconstrained, since all the lensing constraints that were identified were between the two merging subclusters \citep{Mahler2020}. Multiband, high-resolution imaging is necessary for identification of a robust set of lensing constraints, using color and resolved morphology to identify multiple images of the same source.}
 
\done{In this paper, we report new multiband HST imaging of \clustername, obtained to determine whether the system is undergoing a major merger. We also report data and results from the Multi-Unit Spectroscopic Explorer (MUSE) used to measure the redshifts of newly identified lensing constraints, cluster member galaxies, and other sources in the field of view. We present results from an updated lens model based on the new constraints.}

\done{This paper is organized as follows. We describe the HST and MUSE observations, data reduction, and spectra extraction in \autoref{sec:data}. 
We describe the identification and assessment of background lensed galaxies and their spectroscopic confirmation in \autoref{sec:arcs}.
We outline the process used to create our strong gravitational lens model in \autoref{sec:analysis}. We present the lens model results, including a measurement of the masses of the two subcluster components in \autoref{sec:results}.  We discuss our results in  \autoref{sec:disc}. Finally, we summarize our findings in \autoref{sec:conc}.}
    
\done{This work assumes a flat $\Lambda$CDM cosmology with 
$\Omega_{\Lambda} = 0.7$, $\Omega_{m}=0.3$, and $H_0 = 70$ \kms\ Mpc$^{-1}$.
At the redshift of \clustername, $z=\clusterz$, $1\farcs0 = 8.063$ kpc in this cosmology. Magnitudes are reported in the AB system. }

\section{Data} \label{sec:data}

\subsection{Hubble Space Telescope}
\done{\clustername\ was observed in 2021 as part of a joint Chandra-HST program (Chandra Proposal ID \#22800530, 79 ks; HST-GO-16425, 8 orbits, PI: Mahler) with a primary goal to determine whether \clustername\ is a dissociated major merger by constraining the hot gas from X-rays and the dark matter potential from gravitational lensing in the optical. 
We obtained two orbits of HST imaging with the Wide Field Camera 3 (WFC3) on 2021 March 6. Each orbit was split between F110W and F160W, resulting in a total of 2422 s in WFC3-IR/F110W and 2622 s in WFC3-IR/F160W. 
We obtained six orbits of HST imaging with the Advanced Camera for Surveys (ACS) on 2021 May 8-9, with ACS/F606W and ACS/F814W. The ACS observations are a mosaic of three overlapping pointings to achieve a wide field of view and maximal depth at the center. Each pointing was observed for two orbits, and each orbit split between the two filters, reserving the darker half of the orbit (less Earthshine) for F606W. At the core of the cluster (and the strong-lensing region), the ACS mosaic achieved a three-orbit depth (10,584 s total) in each of F814W and F606W. } \done{To the new observations, we coadded one frame of archival ACS/F606W imaging obtained on 2014 June 25 as part of GO-13412 (PI: Schrabback), with an exposure time of 2320~s.}

\begin{figure*}
    \centering
    \includegraphics[width=6 in]{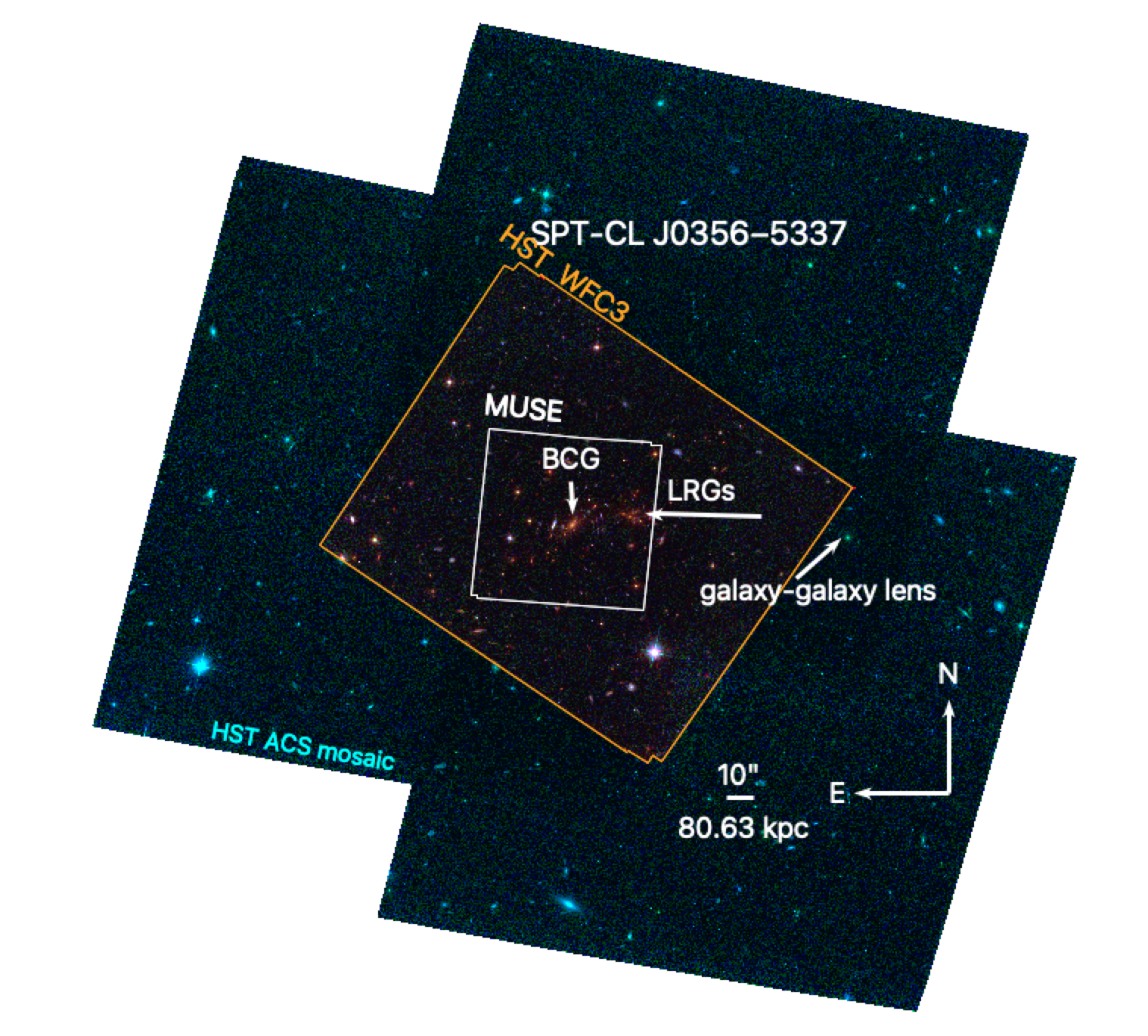}
    \caption{\done{HST imaging of \clustername\ using WFC3/F160W (red), ACS/F814W (green) and ACS/F606W (blue). Cluster components are labeled in white. The different instrument footprints are labeled in white (MUSE), orange (HST WFC3), and cyan (HST ACS mosaic).}}
    \label{fig:fov}
\end{figure*}

\done{The HST data (DOIs: 10.17909/9mak-cy86 ; Titled: HST imaging data for SPT-CL-J0356-5337) were reduced using standard procedures with Drizzlepac\footnote{\url{http://www.stsci.edu/scientific-community/software/drizzlepac.html}}. 
Sub-exposures taken with each filter were combined with \texttt{astrodrizzle}, using a Gaussian kernel with a drop size \texttt{final\_pixfrac}=$0.8$. Images from different visits were aligned using \texttt{tweakreg} and \texttt{tweakback} for a WCS- and pixel-frame matched mosaic, with a pixel scale of $0\farcs03$ per pixel. \autoref{fig:fov} shows the field of view covered with HST. The ACS data cover a field out to $\sim 2\farcm7$ from the cluster center, intended to facilitate weak lensing analysis. The WFC3 data were not tiled, and cover only the cluster core.}

\subsection{MUSE}
\done{We observed the core of \clustername\ with the Multi-Unit Spectrograph Explorer \citep[MUSE;][]{Bacon2010} on the Very Large Telescope (VLT) as part of ESO project 106.21JF.001 (PI: Mahler). The observations were taken between 2021 January 16 and 2021 February 20 for a total exposure time of $14.04$~ks over four different observing blocks, at gray time, with a reported seeing of $0\farcs8$. They were centered on R.A., decl. = [03:56:21.512; $-$53:37:51.954], to maximize the strong lensing region coverage. }

\done{The MUSE data were reduced with the European Southern Observatory (ESO) pipeline \revision{using MPDAF version 3.5 and ZAP version 2.1} \citep{Weilbacher2015} and then reprocessed to improve flat-fielding and sky subtraction using a dedicated pipeline together with MPDAF tools \citep{Bacon2016}, following a methodology similar to the one described in \cite{Fumagalli2016,Fumagalli2017}. We refer the reader to these papers for more details and only give a brief summary of the postprocessing steps here: (1) the individual exposures were resampled to a common astrometric grid defined by the final ESO reduction; (2) a correction of imperfections in the flat fields was applied using the MPDAF self-calibration method as described in \cite{Bacon2017}, and sky subtraction was performed with the ZAP tool, which uses principal component analysis \citep{Soto2016}; and (3) individual exposures were combined to obtain a single data cube using an average $3\sigma$ clipping algorithm. \revision{Finally, we aligned the reduced cube to the HST images.}}

\done{We extracted spectra from the MUSE data following a similar procedure as \cite{Mahler2018}. Spectra were extracted following a mask based on the continuum F814W HST band segmentation map, convolved with a 0\farcs8 FHWM to mimic the effect of the seeing. The redshift identification follows the same classification as \cite{Mahler2018} with maximum confidence, confidence 3, assigned to objects with several clearly identified emission or absorption lines; confidence 2 redshift identifications rely on one emission line (e.g., H$\alpha$, narrow \lya) or a few faint absorption lines (e.g., low signal-to-noise balmer lines); and confidence 1 redshifts are tentative identification and are unreliable but are still recorded. The final catalog contains spectra from 153 sources, of which 32 are classified as cluster member galaxies (see \autoref{sec:selection}) and five were identified as multiply imaged systems (\autoref{fig:constraints}). The spectroscopic redshifts of multiple images used in this work are listed in \autoref{table:constraints} and further discussed in \autoref{sec:constraints}. The rest of the catalog will be published in a later study.}

\done{\subsubsection{Spectroscopic redshifts of Cluster Member Galaxies}
The MUSE field of view is centered on the BCG and provides spectroscopic redshifts of galaxies out to $\sim41\arcsec$ from the BCG. This includes the group of luminous red galaxies (LRGs) projected $\sim21\arcsec$ west of the BCG, which we refer to as the ``LRG core'' or ``LRG subcluster'' in this paper. 
The redshift distribution of the 32 cluster members within the MUSE field of view confirms the findings of \cite{Mahler2020}, that the two main cores of \clustername, the ``BCG core'' and the ``LRG core'', are at the same redshift. Quantitatively, a tight group of seven galaxies in the ``LRG core'' have an average velocity offset from the BCG core of just a few \kms\ (median 135 \kms). We take the redshift of the cluster as the median of 21 galaxies in the cluster core in the range $1.03<z<1.04$, $z_{cluster}=\clusterz$. An elaborate dynamical investigation of this cluster will be presented in a forthcoming publication (G. Mahler et al., in preparation).}

\startlongtable
\begin{table*}[h!]
\begin{center}
\begin{tabular}{c c c c c c c }
 \hline
 ID & R.A. (J2000) & decl. (J2000) & \textit{z} & rms [\arcsec] & $\mu_{median}$ & $\mu_{best}$ \\
 \hline
  1.1  & 59.0852022 & -53.6314475 & 2.363 & 0.22 & $8.5_{-1.5}^{+3.1}$ & $8.5$ \\
  1.2  & 59.0853168 & -53.6307905 & 2.363 & 0.03 & $6.1_{-1.2}^{+2.0}$ & $6.0$ \\
  1.3  & 59.0887924 & -53.6272615 & 2.363 & 0.30 & $2.5_{-0.3}^{+0.6}$ & $2.5$  \\
  11.1 & 59.0852061 & -53.6317905 & 2.363 & 0.38 & $4.9_{-0.7}^{+1.6}$ & $5.0$ \\
  11.2 & 59.0855040 & -53.6303465 & 2.363 & 0.30 & $5.7_{-1.3}^{+3.5}$ & $6.0$ \\
  11.3 & 59.0887949 & -53.6273977 & 2.363 & 0.36 & $2.7_{-0.3}^{+0.7}$ & $2.6$ \\
  2.1  & 59.0843011 & -53.6316673 & 2.364 & 0.13 & $4.7_{-0.6}^{+1.5}$ & $4.7$ \\
  2.2  & 59.0846952 & -53.6300065 & 2.364 & 0.09 & $6.7_{-1.6}^{+4.8}$ & $7.2$ \\
  2.3  & 59.0872684 & -53.6272840 & 2.364 & 0.12 & $2.9_{-0.4}^{+0.7}$ & $2.9$ \\
  21.1 & 59.0842829 & -53.6315221 & 2.364 & 0.05 & $5.6_{-0.8}^{+1.9}$ & $5.6$ \\
  21.2 & 59.0846032 & -53.6301888 & 2.364 & 0.08 & $7.3_{-1.7}^{+4.7}$ & $7.8$ \\
  22.1 & 59.0842691 & -53.6314058 & 2.364 & 0.09 & $6.7_{-1.1}^{+2.3}$ & $6.7$  \\
  22.2 & 59.0845130 & -53.6303539 & 2.364 & 0.14 & $8.5_{-1.9}^{+4.8}$ & $9.1$ \\
  3.1  & 59.0828569 & -53.6331037 & 3.048 & 0.13 & $2.5_{-0.3}^{+0.9}$ & $2.6$  \\
  3.2  & 59.0838084 & -53.6289998 & 3.048 & 0.04 & $3.2_{-0.6}^{+1.0}$ & $3.1$  \\
  3.3  & 59.0856499 & -53.6270832 & 3.048 & 0.13 & $4.4_{-0.8}^{+1.7}$ & $4.2$ \\
  31.1 & 59.0827714 & -53.6330331 & 3.048 & 0.12 & $2.6_{-0.3}^{+0.9}$ & $2.6$ \\
  31.2 & 59.0836965 & -53.6290506 & 3.048 & 0.02 & $2.7_{-0.5}^{+0.9}$ & $2.5$ \\
  31.3 & 59.0855083 & -53.6270606 & 3.048 & 0.14 & $4.4_{-0.8}^{+1.7}$ & $4.2$ \\
  32.1 & 59.0827509 & -53.6329767 & 3.048 & 0.15 & $2.6_{-0.3}^{+0.9}$ & $2.6$  \\
  32.2 & 59.0836564 & -53.6291151 & 3.048 & 0.02 & $2.3_{-0.4}^{+0.8}$ & $2.1$ \\
  32.3 & 59.0855392 & -53.6269909 & 3.048 & 0.14 & $4.1_{-0.7}^{+1.5}$ & $4.0$ \\
  33.1 & 59.0831075 & -53.6332247 & $3.06_{-0.11}^{+0.12}$ & 0.05 &   $2.5_{-0.3}^{+0.8}$ & $2.5$ \\
  33.2 & 59.0840906 & -53.6289688 & $3.06_{-0.11}^{+0.12}$ & 0.08 &   $3.1_{-0.6}^{+1.1}$ & $3.0$ \\
  33.3 & 59.0860422 & -53.6271129 & $3.06_{-0.11}^{+0.12}$ & 0.20 &   $4.2_{-0.7}^{+1.5}$ & $4.1$ \\
  4.1  & 59.0765359 & -53.6311011 & 3.0205 & 0.08 & $7.2_{-2.9}^{+6.3}$ & $8.4$ \\
  4.2  & 59.0764350 & -53.6303685 & 3.0205 & 0.07 & $11.3_{-8.1}^{+20.6}$ & $16.6$ \\
  4.3c & 59.0766660 & -53.6291637 & 3.0205 & \nodata & $12.3_{-5.6}^{+167.7}$ & $25.3$ \\
  41.1  & 59.0766346 & -53.6311878 & 3.0205 & 0.29 & 
  $6.3_{-2.3}^{+4.4}$ & $7.0$ \\
  41.2  & 59.0763229 & -53.6302903 & 3.0205 & 0.15 & $14.5_{-9.9}^{+159.9}$ & $82.8$ \\
  41.3c & 59.0765992 & -53.6292369 & 3.0205 & \nodata &  
  $12.0_{-5.5}^{+152.3}$ & $22.2$ \\
  5.1c  & 59.0954663 & -53.6321227 & $\sim 2.34$ & \nodata &  $48_{-33}^{+1000}$ & $32$ \\
  5.2c  & 59.0952147 & -53.6326267 & $\sim 2.34$ & \nodata &   $30_{-21}^{+680}$ & $136$ \\
  5.3c  & 59.0944939 & -53.6332017 & $\sim 2.34$ & \nodata &    $40_{-28}^{+940}$ & $58$ \\
  6.1c  & 59.0859156 & -53.6317767 & $3.13_{-0.25}^{+0.33}$ & \nodata & $3.3_{-0.5}^{+1.2}$ & $3.2$ \\
  6.2c  & 59.0859224 & -53.6312904 & $3.13_{-0.25}^{+0.33}$ & \nodata &   $18.5_{-5.6}^{+7.7}$ & $19.1$ \\
  7.1  & 59.0989640 & -53.6297059 & 5.3288 & 0.05 &  $2.8_{-0.4}^{+1.3}$ & $2.9$ \\
  7.2  & 59.0960027 & -53.6342447 & 5.3288 & 0.10 &  $1.5_{-0.2}^{+0.6}$ & $1.4$ \\
  7.3  & 59.0939978 & -53.6358341 & 5.3288 & 0.13 &  $3.9_{-0.6}^{+1.2}$ & $4.0$ \\
  9.1c & 59.0423509 & -53.6318281 & \nodata & \nodata &\nodata&\nodata\\
  9.2c & 59.0418968 & -53.6318360 & \nodata & \nodata &\nodata&\nodata\\
  9.3c & 59.0414829 & -53.6323020 & \nodata & \nodata &\nodata&\nodata\\

\end{tabular}
\caption{Lensing constraints and candidate lensed systems. ID refers to the name of each lensed image.  A `c' after the image ID denotes that the image is a candidate and was not used as a constraint in the model.  The ``$z$'' column tabulates the redshift of each source. Redshifts without error bars are spectroscopically confirmed and were used as hard constraints in the model, while those with error bars are model predictions. 
The spectroscopic redshifts of sources 1, 2, and 3 are from \cite{Mahler2020}.
The redshift of system 33 was included in the optimization as a free parameter. 
The spectroscopic redshifts for sources 4 and 7 were found from the spectral analysis of the MUSE cube (see \autoref{sec:constraints}).
For candidate 5, the model-predicted redshift is uncertain and represents a redshift for which the model produces multiple images at this location, also resulting in a poorly-constrained high magnification.
The redshift of candidate 6 is estimated from the best fit parameters of model B.  
``rms'' represents the scatter in the image plane between the predicted and observed images of each source. The last two columns list the magnification measured at the location of the listed coordinates, with 95\% confidence uncertainties measured from Model B (\autoref{table:best_fit_model}). $\mu_{median}$ is the median from the full MCMC chain; $\mu_{best}$ is measured from the best-fit model.}
\label{table:constraints}
\end{center}
\end{table*}

\section{Lensed Sources}\label{sec:arcs}
\subsection{Identification of Multiple Images} \label{sec:constraints} 
\cite{Mahler2020} identified and spectroscopically confirmed three multiply-imaged lensed sources, all with images appearing between the two subhalos of the cluster. \revision{The spectroscopic identification was part of the follow-up campaign within the SPT collaboration using Magellan/LDSS3, Magellan/FIRE, and the Gemini/GMOS-South spectrograph.} Each source has three images, for a total of nine. They used the single-band HST imaging that was available at the time to use the substructure of each source as an individual constraint, multiplexing two of the systems into three sets of constraints each. We adopt the ID scheme of \cite{Mahler2020} for consistency; these systems are identified as \revision{system 1, source 2, and source 3. The latter was confirmed again with MUSE spectroscopic data using the \Lya\ line, other systems were in the so-called \emph{redshift desert} and did not yield any confirmation. For system 3, we kept the redshift estimated from [OIII] as reported in \cite{Mahler2020} as it is a better tracer than \lya\ of the systemic redshift.}  

\begin{figure*}
    \centering
    \includegraphics[width=6 in]{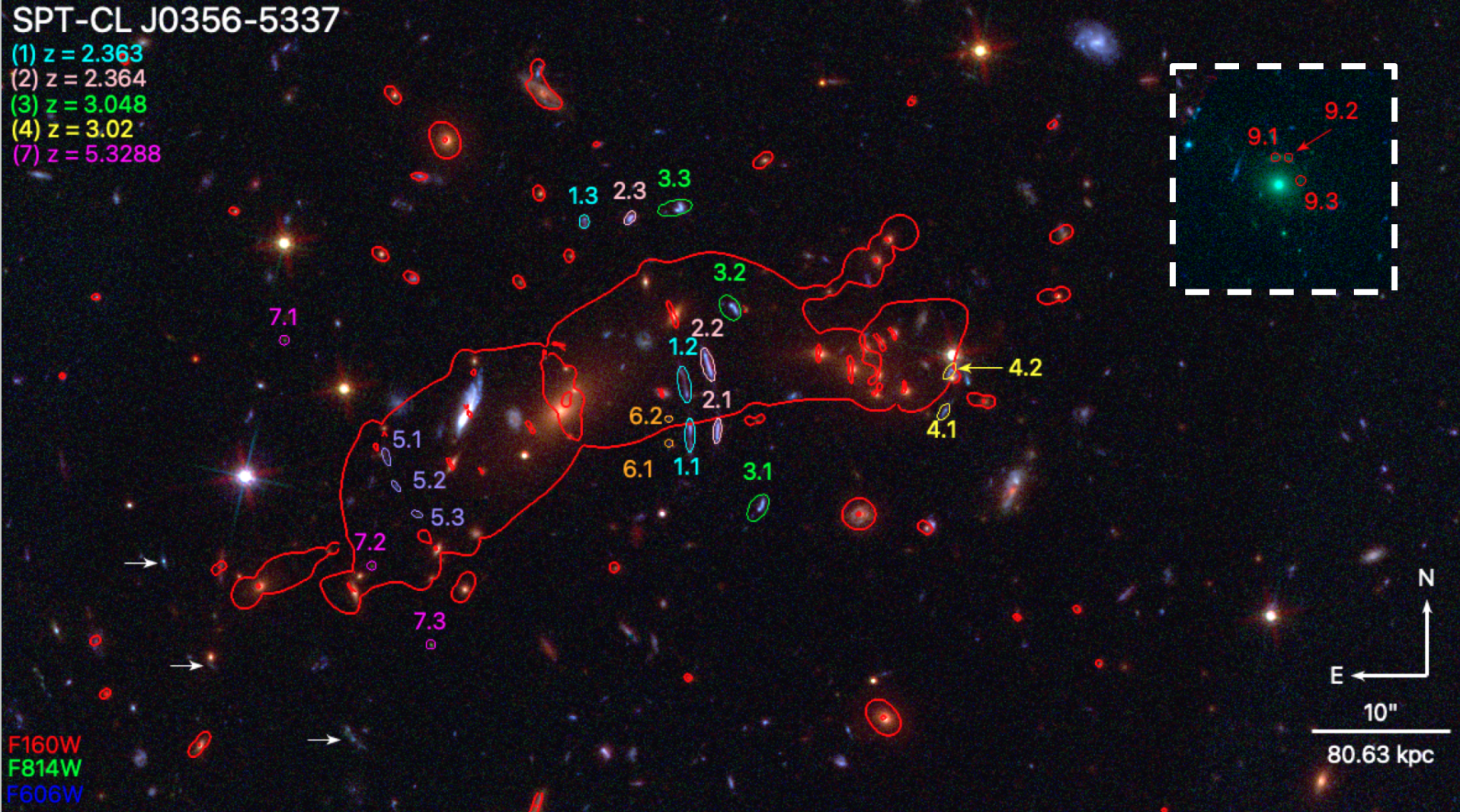}
    \caption{\done{HST image of \clustername\ with lens model constraints labeled. The critical curve at $z=3$ is shown for model B. Arcs and arc candidates are labeled and color-coded by system; their coordinates are listed in \autoref{table:constraints}. White arrows point to the LAEs discussed in \autoref{sec:lya}. An inset shows a galaxy-galaxy lens found $101\farcs1$ west of the BCG (see \autoref{sec:constraints} and \autoref{fig:fov}) at the same scale. The area containing the galaxy-galaxy lens is only covered by ACS and is missing the WFC3-IR filters.}}
    \label{fig:constraints}
\end{figure*}

\done{Using the new HST data, we confirmed the identification of the three secure lensed sources in \cite{Mahler2020} and identified additional lensed galaxies in new, previously unconstrained areas of the field. To the west of the LRGs, we identified system~4, with two distinct emission knots, creating families 4 and 41. We extracted a redshift of $z=\zsystemIV$  for this system from the MUSE data, from a \lya\ emission line (see \autoref{sec:lya}). }

\done{To the east of the BCG, we considered as candidates families 5, 7 and 8.  
Family 7 appears as three images $\sim18\arcsec$ southeast of the BCG. The HST colors as well as the wide separation of $24\farcs5$ between 7.1 and 7.3 point to a lensed source at a relatively high redshift. We spectroscopically confirmed family 7 using the MUSE data at $z=\zsystemVII$, from a \lya\ emission line (see \autoref{sec:lya}).  
Family 5 appears as three faint images with similar colors and geometry consistent with the lensing expectation. However, since we were unable to confirm this source with MUSE, it remains a candidate and was not used as a constraint in the lens model. The lens model places this system at $z\sim2.34$.
Candidate 8, at R.A., decl. $= [59.1011420, -53.6342524]$,  was considered because of its arc-like morphology. However, it was not confirmed as a lensed source: the orientation and lack of obvious counter images make it an unlikely multiply-imaged lensed source that could form a useful constraint. We were not able to determine the redshift of this galaxy from MUSE; its colors are consistent with the cluster red sequence. This galaxy was left out of the analysis.}

\done{Between the two cluster cores, we identified candidate system 6 between the BCG and system 1. It has two arcs straddling the critical curve near 1.2 and 1.3, with blue emission and a lensing geometry similar to that of system 1. No obvious redshift was found in the MUSE data for this family, and given its proximity to system~1 we did not use it to constrain the model despite it being a high-confidence candidate. The model predicts $z\sim3.13$, and a third image $\sim 2$~mag fainter $\sim17$\arcsec\ north of the BCG, where several faint blue sources can be observed.
Finally, we identified a star forming clump near each of the images of system 3, with a similar lensing configuration. We label this family as system 33. As the \Lya\ emission encompasses a wide region around the HST-identified clumps (see \autoref{sec:lya}), we could not extract a unique spectroscopic redshift for this knot to definitively determine whether it is part of source 3. We therefore left its redshift as a free parameter. The lens model-predicted redshift of this source is in agreement with that of source 3. }

\done{We also identified galaxy-galaxy lensing evidence appearing as a partial Einstein ring around an elliptical galaxy, located $101\farcs1$ west of the BCG (system 9; see inset in \autoref{fig:constraints}). We do not have full color data in this area and we were unable to obtain a spectroscopic redshift for the arc, as it is outside of both the WFC3 and MUSE footprints. This arc was not used as a modeling constraint in the final model (see \autoref{sec:halos} for discussion).}

\done{A likely background group of three \lya-emitting (LAE) galaxies at $z=\zsystemX$ was also considered as candidate multiply-imaged system, and ruled out. We describe this group in \autoref{sec:lya}.}

\done{We examined and ruled out the three \textit{candidate multiply imaged systems} proposed by \citet[][their Table 3]{Mahler2020}, based on the inconsistent HST colors and morphology or MUSE spectroscopy. Our candidate image 5.2c coincides with one of the candidates of \citet[][candidate 5.2]{Mahler2020}, however with different candidate counter images. Candidate System~6 in \cite{Mahler2020} coincides with one of the LAE galaxies at \zsystemX\ discussed in \autoref{sec:lya}, and is not multiply imaged.}

\done{\autoref{table:constraints} lists the ID, coordinates, and redshifts of the identified images of lensed sources and candidates. The images are labeled as A.X where A is a number indicating the lensed source ID and X indicates a specific lensed image of source A. For example, source 7 is visible as three lensed images: 7.1, 7.2, and 7.3. We can increase the number of constraints by mapping individual star forming clumps within each image to their counter-images. In this case, we label them as AB.X, where B refers to a single emission knot in system A. For example, system 3 has three distinct emission knots. We refer to these as 31, 32, and 33. Each of these are imaged multiple times, leading to, e.g., 31.1, 31.2, 31.3.  For all constraints used in this lensing analysis, as well as arc candidates, see \autoref{fig:constraints} and \autoref{table:constraints}.}

\subsection{Detection of Extended \Lya\ Emission from Background Galaxies}\label{sec:lya}
\begin{figure}
    \centering
    \includegraphics[width=2.7 in]{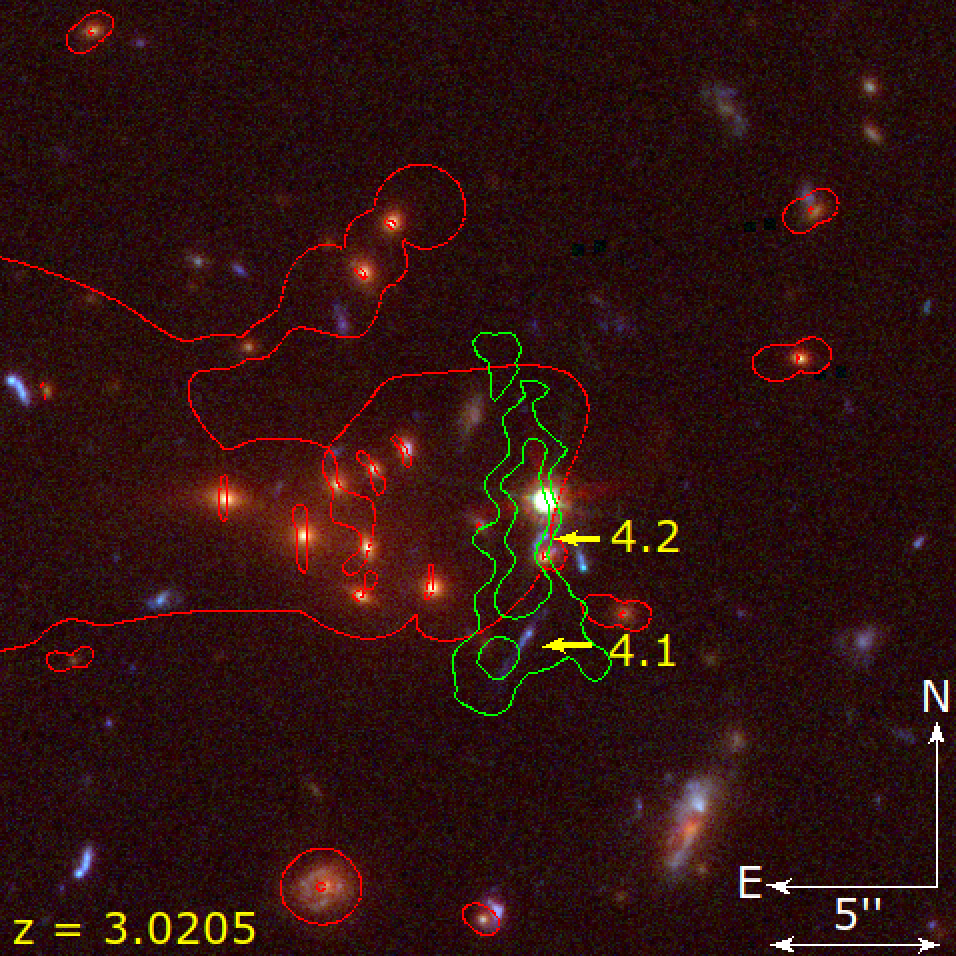}
    \includegraphics[width=2.7 in]{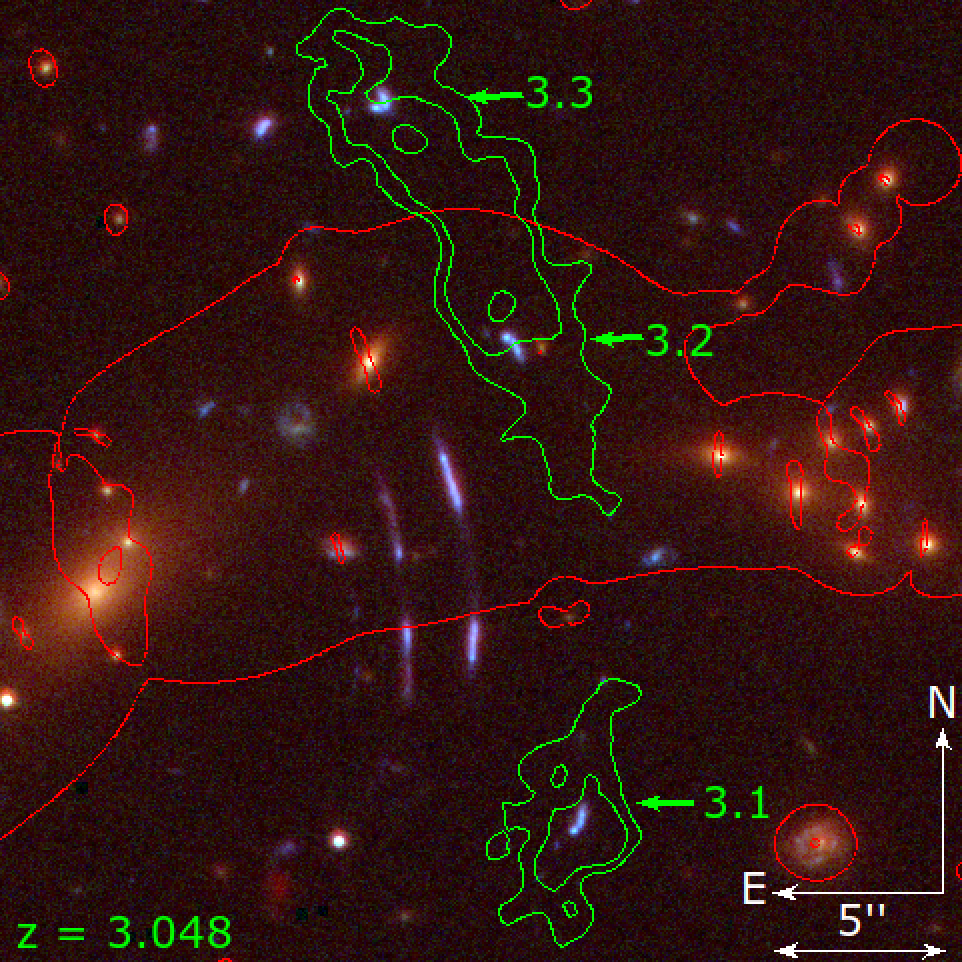}
    \includegraphics[width=2.7 in]{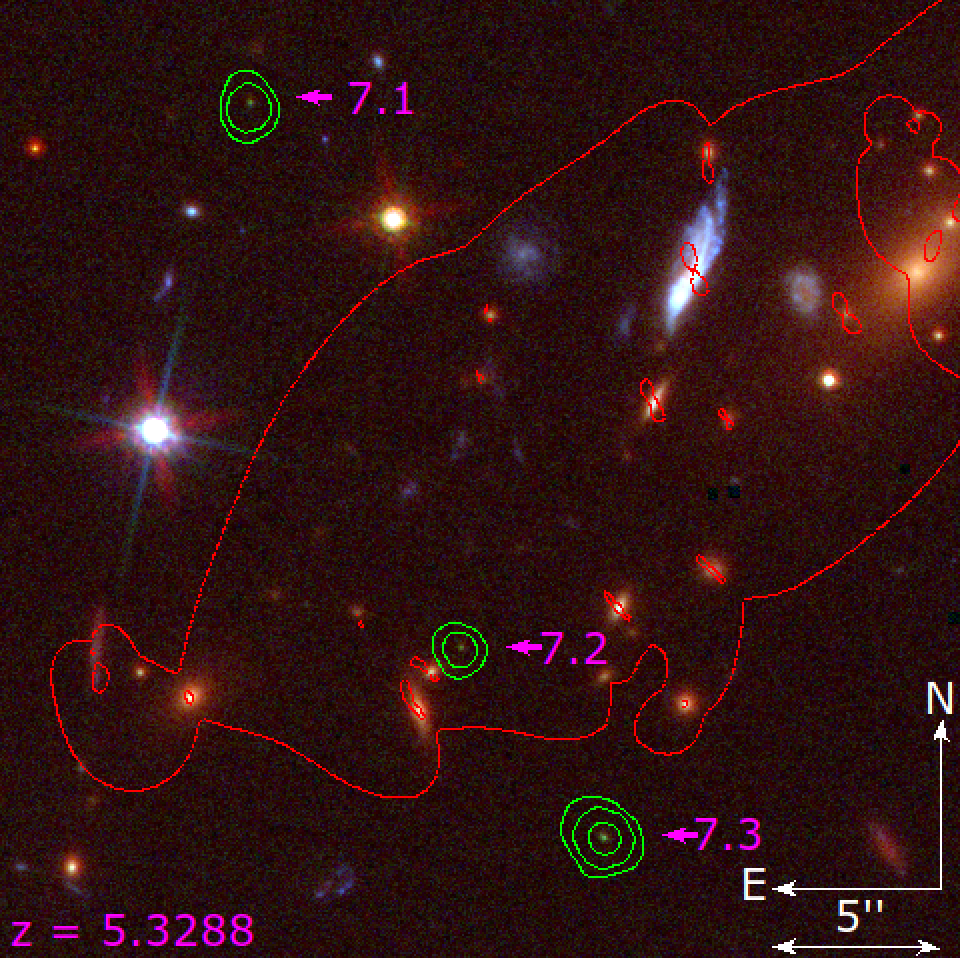}
    \caption{\done{Zoom-in on three strongly lensed galaxies with multiple images, showing \lya\ emission in the MUSE Integral Field Unit (IFU). The \lya\ emission is represented with green contours overplotted on the HST imaging. The top two systems show extended \lya\ emission with peaked emission offset from the HST continuum. The bottom row shows three images of a very compact \lya\ emitter. Contour levels are at {250, 500, 2500, and 5000} \rerevision{$\times 10^{-20}{\rm ergs}~  s^{-1} cm^{-2} $  $arcsec^{-2}$}. The critical curve from the best-fit model is overplotted in red in each panel at the redshift of the emitter.}}
    \label{fig:LAEs}
\end{figure}

\done{Our MUSE observations (see \autoref{sec:constraints}) revealed extended \Lya\ emission associated with several background sources. System~7 was identified in the MUSE data as a new strongly-lensed system, with three images of an LAE galaxy, presenting compact \Lya\ emission at $z=\zsystemVII$, with matching optical counterparts in the HST imaging. System~3 (at $z=\zsystemIII$) and System~4 (at $z=\zsystemIV$) present large diffuse \Lya\ nebulae, where the peaks of the emission are clearly offset from the HST continuum emission. The \lya\ emission from sources 4, 3, and 7 is shown as contours on top of the HST image in \autoref{fig:LAEs}. The magnifications of these arcs are listed in \autoref{table:constraints}. }

\begin{figure*}
    \centering
    \includegraphics[width=\linewidth]{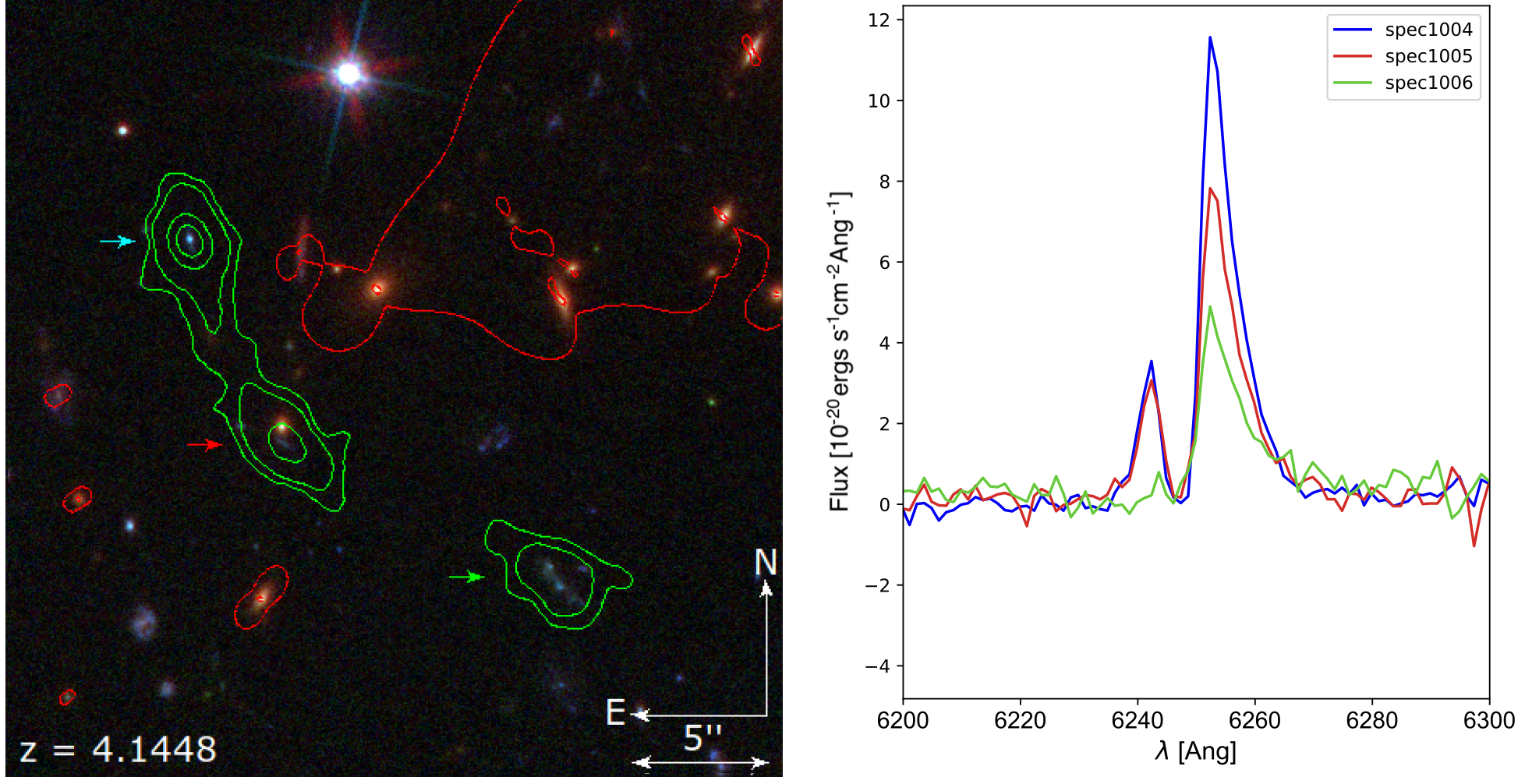}
    \caption{\done{\textit{Left:} HST imaging and \Lya\ contours of three LAEs $\sim30\arcsec$ southeast of the BCG.  
    Contours of the \lya\ levels from MUSE are plotted at \revision{250, 500, 2500, and 5000} \rerevision{$\times 10^{-20}{\rm ergs}~  s^{-1} cm^{-2}$ $ arcsec ^{-2}$}. The critical curve for the redshift of these galaxies is overplotted as a solid red line. 
    \textit{Right:} the extracted spectra of the three sources shown in line colors that match the three arrows in the left panel. The double peak emission is observed at $6254.35$\AA\ and $6254.35$\AA. The redshift of $z=\zsystemX$ is measured from the redder \lya\ peak. Although these three galaxies are at the same redshift, the spectroscopy and imaging data, as well as the lensing analysis, indicate that they are most likely not multiple images of the same source. }}
    \label{fig:nolens-LAE}
\end{figure*}

\begin{figure}
    \centering
    \includegraphics[width=\linewidth]{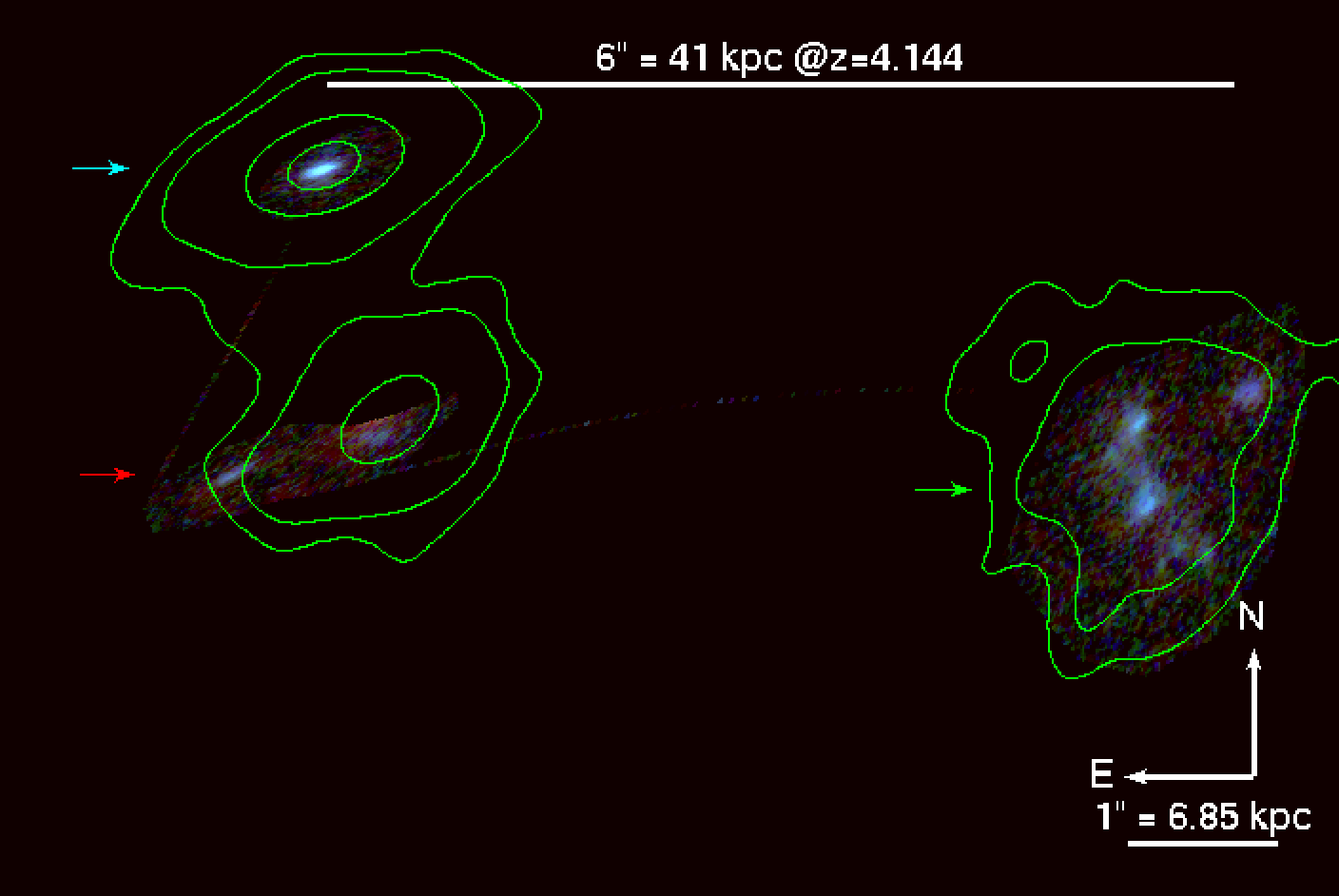}
    \caption{The three LAEs at $z=\zsystemX$, ray-traced to the source plane (\Lya\ contours from MUSE overplotted on the HST imaging). The galaxies are separated by $\sim40$ kpc when unlensed and likely form a small background group.}
    \label{fig:LAE-sourceplane}
\end{figure}

\done{In addition to the secure strongly-lensed LAEs, we have identified double-peaked \Lya\ emission at $z=\zsystemX$ from three sources in the eastern outskirts of the strong lensing regime. \autoref{fig:nolens-LAE} shows the \Lya\ contours overplotted on the HST imaging (left panel), and the double-peaked \Lya\ line of the three galaxies (right panel). \revision{There is no evidence of other nebular lines such as CIV, SiII, FeII, CIII] etc. in the spectra.}
The redshift is assumed from the red \lya\ peak, which is observed at $6254.35$\AA. The blue peak is observed at $6243.80$\AA.  The spatial geometry of the \lya\ emission suggests a multiply-imaged system. However, further examination of this emission leads us to conclude that these are likely not three images of the same galaxy, and may instead be a small background group at $z=\zsystemX$. First, all three \Lya\ detections have an associated optical continuum counterpart in the HST imaging but with strikingly different morphology and inconsistent optical colors and surface brightnesses. Second, while the LAE spectra show very similar redshifts, their blue-to-total flux ratios are  dissimilar among the three galaxies. Finally, the lensing configuration is inconsistent with the constraint on the lensing potential placed by the nearby high-confidence System~7. 
Ray-tracing the three sources through the lens model to their source redshift $z=\zsystemX$, we find that two of the galaxies are separated by approximately 14 kpc in projection, with the third approximately 40 kpc away. 
\autoref{fig:LAE-sourceplane} shows a reconstruction of the $z=\zsystemX$ source plane, done by ray-tracing the three LAEs through the best-fit lens model. We note that choosing a systemic redshift based on the bluer peak or central wavelength has insignificant implication to the lensing interpretation.  
The magnifications of these three LAEs are \muLAEnorth\ for the northern galaxy, \muLAEmiddle\ for the middle galaxy, and \muLAEsouth\ for the southern galaxy, all estimated from} {our fiducial lens model (model B, see \autoref{sec:halos})}. 
An investigation of the circumgalactic gas surrounding this group is beyond the scope of this paper.


\section{Lensing Analysis} \label{sec:analysis}

\subsection{Methodology \label{subsec: method}}
\done{We modeled the cluster using the publicly available parametric lens modeling algorithm \Lenstool\ \citep{Jullo2007}. \Lenstool\ uses Markov Chain Monte Carlo (MCMC) formalism to explore the parameter space and identify the set of parameters that produce the smallest scatter between observed and predicted strong lensing constraints.
As constraints for the lens model, we used the positions of multiple images of lensed sources (\autoref{sec:constraints}) as well as their spectroscopic redshifts when available. 
\Lenstool\ creates models that are linear combinations of galaxy- and cluster-scale halos. In this analysis, all halos were represented with the pseudo-isothermal ellipsoidal mass distribution (dPIE, also referred to in the literature as PIEMD), which has seven  parameters: position $(x, y)$, ellipticity $e$, position angle
$\theta$, core radius $r_{core}$, truncation radius $r_{cut}$, and normalization $\sigma_0$. For cluster-scale halos, we fixed $r_{cut}$ at 1500 kpc, since this parameter is far outside the strong lensing region where lensing constraints are observed; we allowed all other parameters to be optimized by \Lenstool, unless otherwise noted. For galaxy-scale halos, the positional parameters $(x, y, e, \theta)$ were  fixed to the observed positions of their respective cluster member galaxies, as measured with \texttt{Source Extractor} \citep{Bertin1996}, and the other parameters scaled to their magnitudes through scaling relations \citep{Jullo2007} as an ensemble adding only two free parameters to the optimization process.}

\done{Our starting point for the lens model is the model published by \cite{Mahler2020}. With the multiband HST data and this preliminary lens model, we secured four new lensed sources and measured the redshifts of two of these using MUSE data (see \autoref{sec:constraints}). We then converged on the best lens model by iteratively adding complexity to the model. 
The final lens model has positional constraints from 32 multiple images of five spectroscopically confirmed lensed sources and substructure within them.}

\subsection{Selection of Cluster Galaxies} \label{sec:selection}
\begin{figure}
    \centering
    \includegraphics[width=1.0\linewidth]{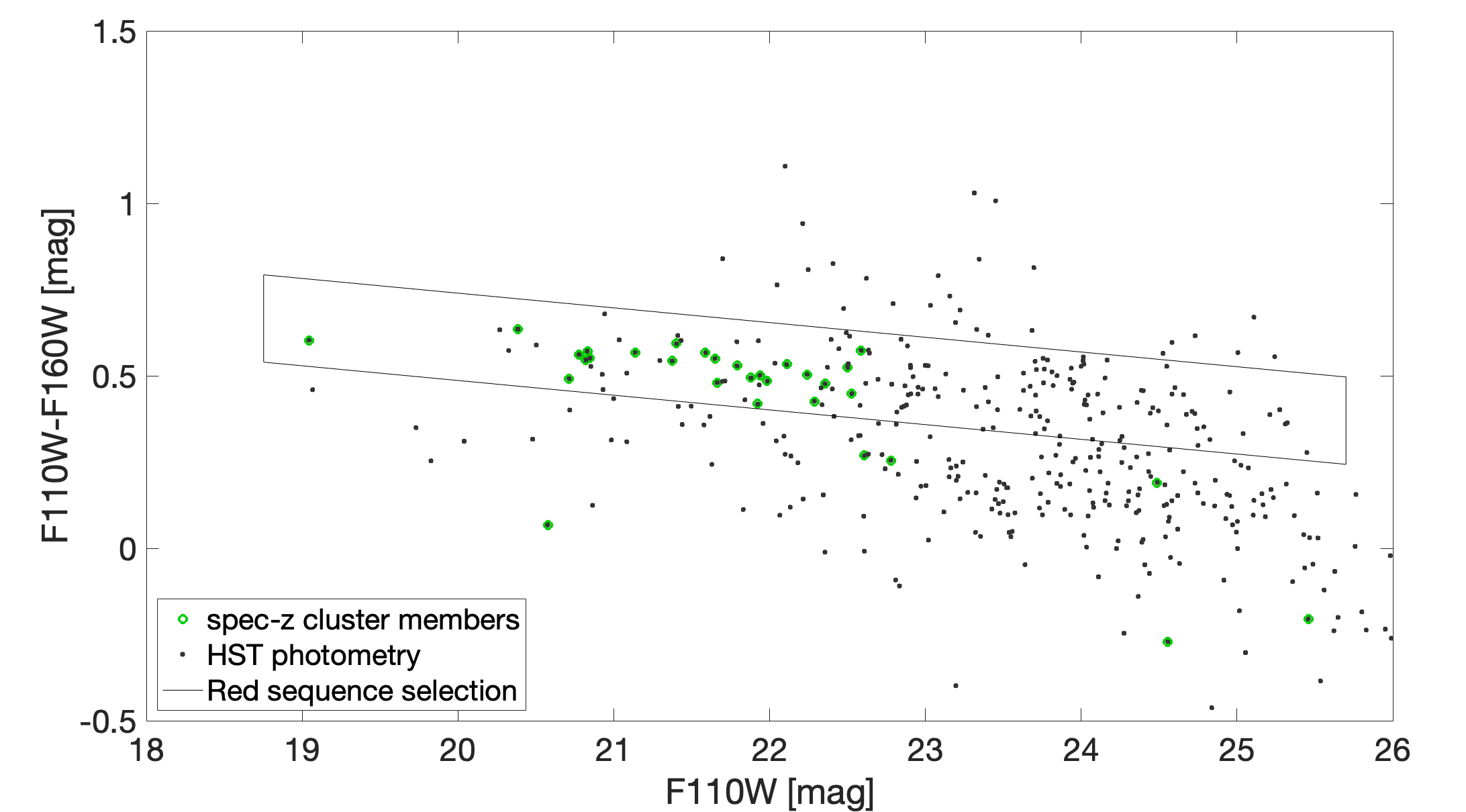}
    \caption{\done{Color-magnitude diagram of galaxies in the HST/WFC3 field of view. The F110W-F160W color is plotted against F110W magnitude. Cluster member galaxies form a red sequence in this parameter space, highlighted by spectroscopically-selected cluster members (green datapoints). The solid line marks the cluster member selection, based on a linear fit to the spectroscopically-selected members.}}
    \label{fig:cmd}
\end{figure}

\done{Cluster member galaxies were selected photometrically and spectroscopically. We generated a photometric catalog of extended sources (galaxies) in the HST/WFC3 field of view using \texttt{Source Extractor} \citep{Bertin1996} in dual-image mode with the F110W image as the detection image, and magnitudes measured within the same apertures in F110W and F160W. We then matched the photometric catalog with the MUSE spectroscopic redshift list by coordinates. \autoref{fig:cmd} shows the F110W$-$F160W vs.\ F110W color-magnitude diagram of galaxies in this field of view. Galaxies with spectroscopic redshifts in the range $1.015<z<1.055$, overplotted in green, highlight the cluster red sequence \citep{gladdersyee2000}. We fit the 32 spectroscopically-confirmed cluster members in the MUSE field of view with a linear fit, using iterative 3$\sigma$ clipping, to define the red sequence and its width in color-magnitude space. We then applied the red-sequence selection (solid line in \autoref{fig:cmd}) to the full photometric catalog to expand the selection to galaxies as faint as F110W=27.5 mag and galaxies outside the MUSE field of view. The final catalog has a total of 149 cluster member galaxies.}

\subsection{Mass Halos} \label{sec:halos}
\begin{figure}
    \centering
    \includegraphics[width=3.25 in]{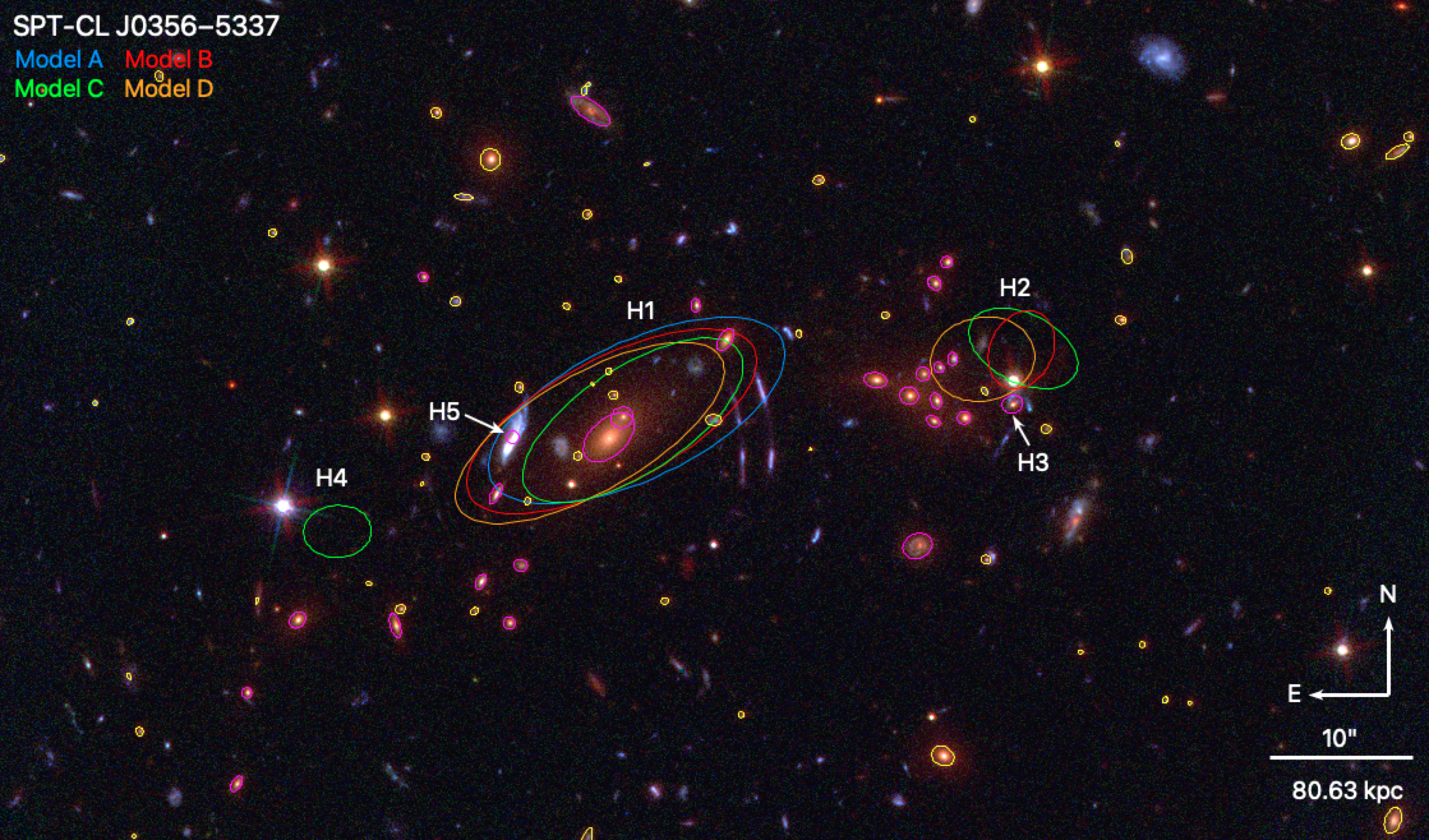}
    \caption{\done{Image of the field with cluster member galaxies identified. Magenta ellipses denote cluster member galaxies spectroscopically confirmed using MUSE data. Yellow ellipses denote cluster member galaxies identified by color (see \autoref{sec:selection}). The dark matter halos described in \autoref{sec:halos} are identified in blue, red, green, and orange for models A, B, C, and D, respectively.}}
    \label{fig:halos}
\end{figure}

\done{Our lens model describes the lens plane as a linear combination of cluster-scale halos and galaxy-scale halos (\autoref{fig:halos}). The cluster member distribution motivates at least two cluster components. We explored the possibility of a more or less complex mass distribution by allowing a third cluster-scale halo or removing the second cluster-scale halo. This is further discussed in \autoref{sec:disc}. As noted above, all the halos are parameterized as dPIE.
Halo 1 (H1) is a cluster-scale dark matter halo associated with the BCG sub-cluster. H2 is a cluster-scale dark matter halo associated with the LRG subcluster. All parameters of these two halos were allowed to vary, with the exception of $r_{cut}$ as explained in \autoref{subsec: method}. H3 is a galaxy-scale halo, centered on a galaxy observed near the images of system 4; we freed the slope and normalization parameters of this galaxy, to better account for small disturbances in the lensing potential in this region, while keeping $x$, $y$, $e$, and $\theta$ fixed to their observed properties as measured by \texttt{Source Extractor}. H4 was added to the model east of system 7, to explore whether additional mass in that region was needed in order to reproduce the lensing evidence and the apparent high ellipticity of the lens. All the parameters of this halo were allowed to vary within broad priors. }

\done{H5 is a galaxy-scale halo that accounts for the mass of the ``jellyfish'' galaxy \citep{Mahler2020}, a bright infalling star-forming cluster member galaxy $7''$ east of the BCG. This galaxy was not included in the cluster-member catalog, as it does not follow the same mass-luminosity relation as quiescent cluster member galaxies. 
Using the same mass-luminosity relation would have assigned too much mass to this galaxy -- the jellyfish galaxy is apparently star-forming, and is therefore brighter than a similarly-massive elliptical galaxy would be. We included this halo in the model as an independent galaxy-scale halo decoupled from the scaling relations of typical cluster members, restricting its parameters so that its mass would remain consistent with a cluster-member galaxy at this redshift.}

\done{In some iterations of the lens model, we freed the galaxy-galaxy lens halo (observed $101\farcs1$ west of the cluster core), allowing its parameters to be constrained by system 9 (shown in the inset in \autoref{fig:constraints}). Freeing the galaxy-scale halo had no detectable effect on the model, so we ultimately left this galaxy linked to the other galaxies in the galaxy catalog. Consequently, system 9 was not used to constrain the cluster lens model.}
\done{We also investigated whether the BCG should be decoupled from the mass-luminosity relation of the rest of the cluster-member galaxies, by freeing the normalization and slope parameters of the BCG ($\sigma_0$, $r_c$, $r_{cut}$). In all the tested cases, the models converged on similar results as models with these three parameters linked to the scaling relations and did not produce better rms or $\chi^2$. Our models therefore treat the BCG the same as the rest of the cluster members. }

\done{In total, the models have 149 galaxy scale halos, and 1-3 cluster-scale halos. }

\begin{table*}
\begin{center}
\begin{tabular}{l c c c c c c c c}
\hline
Model & Halo & $\Delta$ R.A. & $\Delta$ decl. & \textit{$e$} & \textit{$\theta$} & \textit{$\sigma_{0}$}\ & \textit{r$_{cut}$}  & \textit{r$_{c}$} \\
 &  & [arcsec] & [arcsec] &  & [deg]  & [km s$^{-1}$] &[kpc]  &[kpc]\\

\hline
A: H1   & H1 & $-2.05^{+0.85}_{-0.93}$ &$1.94_{-0.51}^{+0.47}$  &$0.74_{-0.10}^{+0.06}$  & $25.4_{-1.1}^{+1.4}$ & $939_{- 46}^{+ 44}$ & [1500] & $ 82_{- 12}^{+ 10}$ \\
one cluster halo      & H3 & $[-28.31]$ & [2.34] & [0.18] & [19.08] & $ 93_{- 15}^{+ 95}$  & $ 20_{- 19}^{+ 74}$ & $0.06_{-0.04}^{+0.43}$ \\
rms=$0\farcs20$, $\chi^2=14.03$, k=$15$ & H5 & [6.69] & [$-0.03$] & [0] & [0] & $ 80_{- 24}^{+164}$ & $ 29_{- 27}^{+ 19}$ & $0.03_{-0.03}^{+0.06}$\\          
$\ln(\mathcal{L})=11.23$, BIC=$32.9$            & L* & \nodata & \nodata  & \nodata & \nodata & $247.7_{-35.5}^{+1.5}$ & $ 53_{- 11}^{+ 25}$ & [0.15] \\
\hline
B: H1+H2  & H1 & $-0.20^{+2.95}_{-0.88}$ & $1.12_{-1.57}^{+0.68}$ & $0.73_{-0.12}^{+0.07}$ & $27.4_{-2.8}^{+2.8}$ & $884_{-166}^{+ 30}$ & [1500] & $ 68_{- 27}^{+ 10}$ \\
two cluster halos      & H2 &$-28.9^{+13.6}_{-1.6}$ & $6.2_{-3.9}^{+3.6}$ &  $0.26_{-0.23}^{+0.60}$ & $243_{-137}^{+ 30}$ & $373_{- 20}^{+186}$ & [1500] & $ 37_{- 11}^{+ 62}$ \\
rms=$0\farcs17$, $\chi^2=9.89$, k=$21$            & H3 & $[-28.31]$ & [2.34] & [0.18] & [19.08] & $ 51_{- 13}^{+142}$ & $ 81_{- 80}^{+ 16}$ & $0.15_{-0.14}^{+0.33}$ \\
$\ln(\mathcal{L})=13.3$, BIC=$50.9$            & H5 & [6.69] & [$-0.03$] & [0] & [0] & $150_{- 96}^{+ 95}$ & $ 14_{- 12}^{+ 35}$ & $0.01_{-0.01}^{+0.08}$\\
            & L* & \nodata & \nodata  & \nodata & \nodata & $246.5_{-51.5}^{+2.7}$ & $43.1_{-6.8}^{+34.7}$ & [0.15] \\
\hline
C: H1+H2+H4  & H1 &$-1.7^{+2.6}_{-1.3}$ & $1.2_{-1.3}^{+1.0}$ & $0.70_{-0.15}^{+0.10}$ & $33.1_{-6.6}^{+7.2}$ & $739_{-200}^{+141}$ & [1500] & $ 44_{- 27}^{+ 25}$ \\
three cluster halos   & H2 & $-29.1^{+13.8}_{-1.8}$ & $6.3_{-2.1}^{+1.9}$ & $0.48_{-0.42}^{+0.36}$ & $154_{- 38}^{+104}$ & $448_{- 86}^{+183}$ & [1500] & $24.4_{-3.2}^{+72.5}$ \\
rms=$0\farcs14$, $\chi^2=6.91$, k=$28$            & H3 & $[-28.31]$ & [2.34] & [0.18] & [19.08] &$40.2_{-9.2}^{+67.2}$ & $ 88_{- 85}^{+ 10}$ & $0.28_{-0.27}^{+0.21}$ \\
$\ln(\mathcal{L})=14.79$, BIC=$73.7$            & H4 & $19.0^{+10.6}_{-4.4}$ &$-6.6_{-6.6}^{+1.4}$& $0.26_{-0.24}^{+0.02}$ & $3.48_{-0.91}^{+173.61}$ & $312_{-195}^{+182}$  & $348_{-206}^{+624}$ & $16.6_{-5.5}^{+66.4}$ \\
            & H5 & [6.69] & [$-0.03$] & [0] & [0] & $81_{- 24}^{+163}$ & $12.1_{-9.6}^{+36.7}$ &$0.07_{-0.06}^{+0.03}$ \\
            & L* & \nodata & \nodata  & \nodata & \nodata & $231_{- 41}^{+ 17}$ & $ 62_{- 28}^{+ 16}$ & [0.15] \\ 
\hline
D: fixed galaxies (1) & H1 & $1.3^{+1.8}_{-1.8}$ & $0.3_{-1.3}^{+1.0}$ & $0.74_{-0.21}^{+0.05}$  & $29.8_{-3.4}^{+2.1}$ & $823_{-82}^{+89}$ & [1500] & $47_{-18}^{+21}$ \\
H1+H2 & H2 & $-26.2^{+10.5}_{-2.8}$ & $5.5_{-2.5}^{+2.4}$ & $0.22_{-0.21}^{+0.22}$ & $190_{-82}^{+81}$ & $494_{-134}^{+109}$ & [1500] & $43_{-18}^{+34}$ \\
rms=$0\farcs17$, $\chi^2=10.62$, k=$19$        & H3 & $[-28.31]$ & $[2.34]$ &$ [0.18]$ & [19.08] & $55_{-14}^{+124}$ & $88.8_{-86.8}^{+5.8}$  & $0.34_{-0.33}^{+0.15}$ \\
$\ln(\mathcal{L})=12.93$, BIC=$44.2$        & H5 & $[6.69]$ & $[-0.03]$ & $[0]$ & $[0]$ & $123_{-69}^{+121}$ & $6.8_{-4.1}^{+41.8}$ & $0.07_{-0.06}^{+0.03}$ \\
            & L* & \nodata & \nodata  & \nodata & \nodata & [217] & [39]  & [0.15] \\
\hline
E: fixed galaxies (2) & H1 & $0.2^{+2.1}_{-2.5}$  & $0.5^{+1.7}_{-1.3}$  & $0.65_{-0.10}^{+0.14}$ & $33.4^{+0.9}_{-9.4}$ & $743^{+151}_{-45}$ & [1500] & $40^{+26}_{-15}$ \\
H1+H2+H4  & H2 & $-20.1^{+4.2}_{-10.7}$  & $6.0^{+1.5}_{-2.4}$  & $0.34^{+0.51}_{-0.32}$ & $175_{-71}^{+99}$ & $508^{+95}_{-131}$ & $[1500]$ & $49^{+30}_{-23}$ \\
rms=$0\farcs15$, $\chi^2=8.3$, k=$26$        & H3 & $[-28.31]$ & $[2.34]$ & $[0.18]$ & $[19.08]$ & $88^{+46}_{-43}$ & $43^{+53}_{-39}$ & $0.35^{+0.14}_{-0.34}$ \\
$\ln(\mathcal{L})=14.09$, BIC=$67.7$        & H4 & $18.6^{+10.6}_{-3.8}$ & $-8.1^{+2.7}_{-11.5}$  & $0.19^{+0.10}_{-0.19}$ & $31^{+144}_{-26}$ & $257^{+133}_{-138}$ & $620^{+370}_{-480}$ & $18^{+80}_{-7}$ \\
        & H5 & [6.69] & [$-0.03$] & [0] & [0] & $179^{+67}_{-123}$ & $41^{+7}_{-39}$ & $0.01_{-0.00}^{+0.09}$\\
            & L* & \nodata & \nodata  & \nodata & \nodata & [217] & [39]  & [0.15] \\
\hline
\end{tabular}
\caption{{Best-fit parameters of lens models of \clusterfullname. Fixed parameters are in brackets. $\Delta$ R.A.\ and $\Delta$ decl.\ are measured relative to the reference coordinate (R.A. = 59.0895910, decl. = $-53.6311098$). Ellipticity $e$ is measured as $(a^2-b^2)/(a^2+b^2)$ where $a$ and $b$ are the semimajor and semiminor axes of the ellipse. $\sigma_0$, $r_{c}$, and $r_{cut}$ are the normalization, core radius, and cut radius of the halo, respectively. Uncertainties represent the 95th percentile from the full MCMC sampling. All the models have the same number of constraints, n=2$\times$(\#arcs $-$ \#sources)=40. k is the number of free parameters. $\chi^2$ and $\ln(\mathcal{L})$ are calculated by \lenstool\ with a uniform positional uncertainty of $0\farcs3$. Cluster-scale halos H1, H2, and H4 are near the BCG, LRG group, and the eastern outskirts, respectively; H3 is a galaxy in the LRG group near system 4; H5 is the jellyfish galaxy. L* gives the pivot parameters of the cluster member scaling relations. See \autoref{sec:halos}} and \autoref{fig:halos}.}
\label{table:best_fit_model}
\end{center}
\end{table*}
\section{Results} \label{sec:results}
\done{The multiband HST imaging and MUSE spectroscopy resulted in the robust identification of lensing constraints in previously underconstrained areas of the lens. Improving upon the previous model, the mass distribution is now better constrained in the center and, most importantly, east of the BCG and west of the LRGs. 
}

\done{As described above, the iterative process of converging on the best description of the lens plane that satisfies the lensing constraints starts with a simple model and gradually increases the complexity until new models do not significantly improve the goodness of fit. }\done{We investigated several lens models with increasing complexity levels and numbers of free parameters spanning a range of modeling choices. Model A, with the least number of constraints, describes the lens plane with one cluster-scale halo (H1) plus cluster-member galaxies. Model B adds a second halo (H2) near the group of LRGs. Model C adds a third halo (H4) in the eastern part of the cluster. We also present results from models D and E, which are similar to models B and C, respectively, except that the scaling relations of cluster-member galaxies are fixed, limiting the galaxy contribution to about $10\%$ of the total cluster mass.}

\done{The parameters of the {five} models are tabulated in \autoref{table:best_fit_model}. 
Statistical uncertainties on the best-fit model parameters were obtained from the MCMC sampling of the parameter space, and represent a 95\% confidence interval. Parameters that were not optimized are listed with square brackets. }

\done{To assess the lens models we evaluated two criteria: the image-plane rms and the Bayesian information criterion (BIC). }
\done{The rms is the average image plane scatter between the observed positions of lensed images and their positions predicted by the model. A model is generally considered better if it has a lower rms.} 
\done{According to this criterion, the favored lens model is the one with the highest flexibility, Model C, which allows three cluster-scale halos (H1, H2, and H4), and solves for the normalization of the cluster members' scaling relation as free parameters. This model achieved an image plane rms of $0\farcs14$.}

\done{The second criterion, BIC, provides a quantitative method of comparing similar models by penalizing a higher number of free parameters, which helps to prevent overfitting. The BIC is calculated using the following equation \citep{Schwarz1978}:}

\begin{equation}
BIC=-2\ln(\mathcal{L})+k\ln(n),
\end{equation}

\noindent{\done{where $\mathcal{L}$ is the likelihood, $k$ is the number of free parameters, and $n$ is the sample size, which is chosen to be the number of constraints, $n$=[2$\times$ the number of multiple images $-$ 2$\times$ the number of sources] (see \citealt{lorah2019} for a discussion of the difference between the sample size and the number of constraints).
A model with lower BIC is considered to be statistically preferred.}}

\done{The BIC values are listed in \autoref{table:best_fit_model}.} 
\done{We find that the BIC increases with the number of parameters. Although models with higher complexity result in better recovery of the lensing evidence, the high penalty of the BIC makes the simpler models statistically favored. 
As we demonstrate in the following sections, some results remain consistent among models, indicating that they are robust against systematic uncertainties related to modeling choices. In particular, these include the masses of the two major subcluster components and their mass ratio, an important measurement for classifying the system as a major merger. Purely based on the BIC, the simplicity of model A is favored as this model shows the lowest BIC of 32.9. We will discuss this in \autoref{sec:disc}.}

\done{In terms of achieving a low rms, models D and E, with fixed galaxy scaling relations, perform as well as their equivalent models with free galaxy scaling relations parameters (B and C, respectively). Formally, these models would be statistically favored over their nonfixed counterparts, as they have two fewer parameters and thus lower BIC. However, since one of the questions we are aiming to answer with the lensing analysis is the fraction of mass contained in galaxies, keeping this component fixed significantly limits our ability to investigate this question. Of these two models, the BIC  favors model D over model E.}

\done{Assessing both criteria, we select model B as the one that best balances an adequate sampling of the parameter space, good reproduction of the lensing constraints, and lower number of free parameters. We discuss this selection in \autoref{sec:disc}. 
In the following sections we quote the results from model B as the best-fit model. We present the outputs of all five lens models in tables, and draw conclusions from comparing and contrasting these models.
To prevent visual clutter, we only include models A, B, C, and D in the figures unless otherwise specified}.  

\subsection{Mass distribution}\label{sec:massresults}
\begin{figure}
    \centering
    \includegraphics[width=3.25 in]{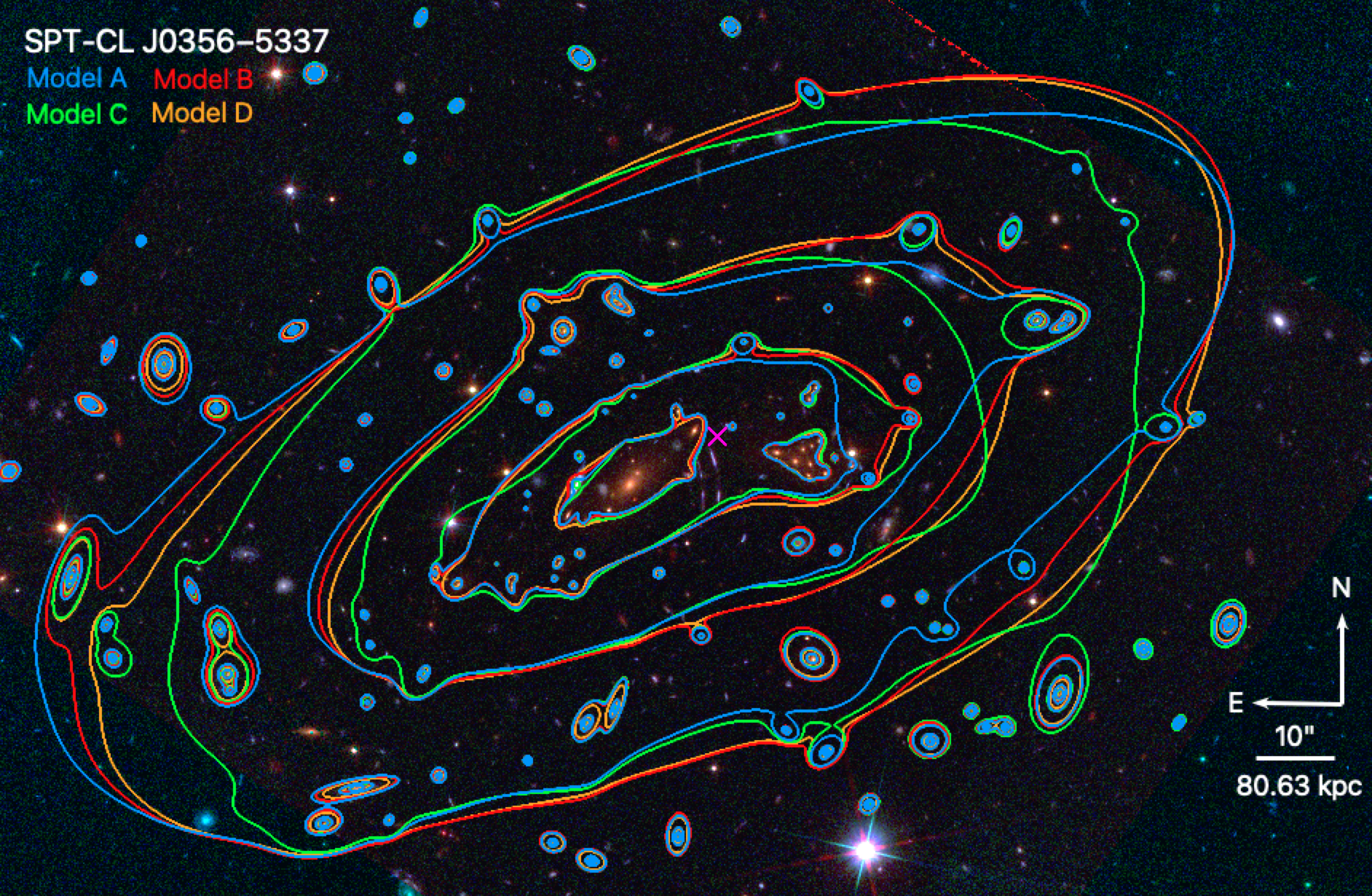}
    \caption{\done{Mass contours overlaid on the HST imaging for models A, B, C, and D in blue, red, green, and orange, respectively. The contours are plotted at projected mass densities of 0.25, 0.5, 1 and 2$\times 10^{9} {{\rm M}_\odot} {\rm kpc}^{-2}$.} \revision{The magenta cross marks the centroid the X-ray emission (Mahler et al. in prep.)}}
    \label{fig:mass-contours}
\end{figure}

\begin{figure*}
    \centering
    \includegraphics[width=1\linewidth]{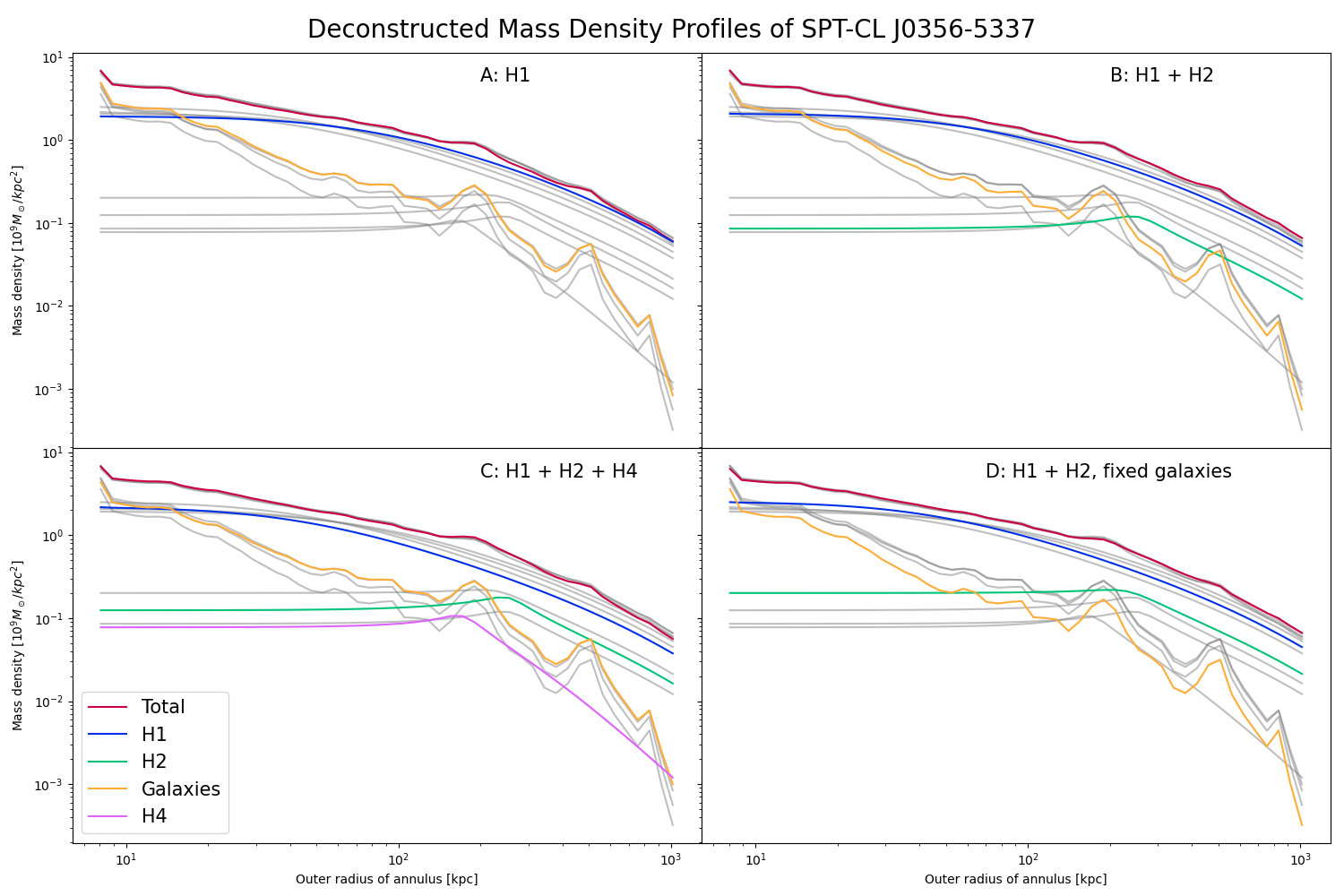}
    \caption{{The projected mass density profile of the individual contributions of the major cluster halos and  the galaxy component. The average mass densities are measured in annuli centered on the BCG. Uncertainties are on the order of $\sim$10\%, but are omitted for clarity. Each panel displays the results from one of the models. The model components are color-coded as indicated in the legend. Gray lines are the same in all panels, and are plotted to facilitate a convenient comparison between the models. }}
    \label{fig:mass-profile}
\end{figure*}

\done{\autoref{fig:mass-contours} shows the mass contours derived from the best-fit lens models {A, B, C, and D}. The projected mass density profile, centered on the BCG, is shown in \autoref{fig:mass-profile}.}

\done{The total projected mass of the cluster, measured within 500~kpc of the BCG, is  
\massclusterB\ (cylindrical mass), derived from {model B} (see \autoref{table:masses} for results from all models). We note that the 95th percentile statistical uncertainties underestimate the true uncertainty at this radius, as the measurement out to 500 kpc extends beyond the strong lensing regime. Regardless, the discovery of new multiple images outside the region between the  two main halos provides constraining power on the location and extent of these halos, leading to a more accurate and precise mass estimate than was previously possible. Overall, the statistical uncertainties are reduced by at least $50\%$ compared to previous work.}

\done{We measured the masses enclosed within projected radius $R$ of the different subcluster components (the ``BCG cluster core'', hereafter $M_{BCG}(<R \text{kpc})$, and the ``LRG core'', hereafter $M_{LRG}(<R \text{kpc})$) by summing the projected mass density output of the strong lensing models inside an aperture with a radius of 80~kpc, centered on each component. The BCG component is centered on the BCG at R.A.\ = $59.0895910$, decl.\ = $-53.6311098$, and the LRGs component is centered on the brightest cluster member galaxy in the group, at R.A.\ = $59.079727$, decl.\ = $-53.630291$. Following \cite{Mahler2020}, we used a projected radius of 80~kpc because it encloses each component without overlapping with the other. We calculated uncertainties using a 95\% confidence interval.}

\begin{figure*}
    \centering
    \includegraphics[width=1\linewidth]{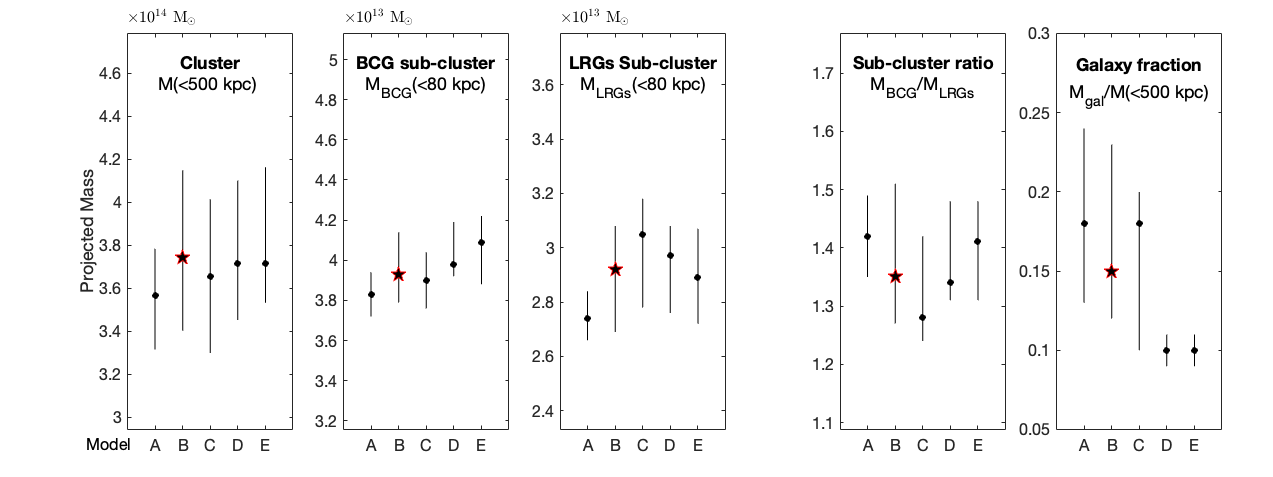}
    \caption{\done{Mass and mass fraction from the different lens models presented in this work. From left to right: the total mass of the cluster measured as projected (cylindrical) mass within 500~kpc from the BCG; the mass measured within 80 kpc from the BCG, the mass measured within 80 kpc from the brightest galaxy in the LRG core, the ratio of the two subclusters; and the mass fraction contained within galaxy-scale halos. The galaxy mass fraction includes the individually optimized galaxies (H3 near system 4, and the jellyfish galaxy H5). The datapoints match the values listed in \autoref{table:masses} and their 95\% confidence interval. The red star symbols highlight our fiducial model (model B). Measured masses and subcluster mass ratio agree among all models within the uncertainties. The galaxy fraction differs between models mainly due to modeling choices; models D and E have a fixed galaxy scaling relation and only allow the cluster-scale potential to vary. Model B results in a galaxy fraction slightly lower than the other models.} }
    \label{fig:masses}
\end{figure*}

\done{We find that the mass of the BCG cluster core is \massBCGB, the mass of the LRG core is \massLRGsB, and their mass ratio is $1:1.35^{+0.16}_{-0.06}$, all from {model B}. We find similar results for the masses and mass ratios for {all the models considered in this work}; they are listed in \autoref{table:masses} and presented in \autoref{fig:masses}.  The statistical uncertainties are estimated from the MCMC sampling of the parameter space. To account for possible correlations between the masses within the model, in calculations of mass ratios, we first calculated the mass ratio in each sampled step in the MCMC and then determined the 95\% uncertainty.   
The implications of this result are discussed in \autoref{sec:disc}.} 

\revision{The parametric nature of the lens model algorithm allows us to consider an alternative approach, by disentangling the contributions from different components and calculating their masses analytically. To estimate the mass ratio of the two main cores, we calculated the mass within} \rerevision{160} 
\revision{kpc of H1 and the BCG and compared it with H2 and the eight brightest LRG in the LRG group using their dPIE parameterization and equations (A10) and (A25) in \cite{Eliastottir07}.}
\rerevision{The projected radius of 160~kpc was chosen as it the largest aperture that is well-constrained by strong lensing.}
\revision{We then compared the mass of H1 plus the BCG to that of H2 plus the 8 LRGs, to obtain the mass ratios of the two main cores, resulting in }
\rerevision{$3.2\pm1.2$}.
\revision{The 95\% uncertainty was derived from repeating the calculation with parameters from each sample step in the MCMC. 
We take the ``deconstructed'' mass estimates and ratios with caution. As we thoroughly discuss in \autoref{subsec:degeneracies}, while the total projected mass is indeed well constrained and robust against modeling choices, model degeneracies mean that the mass of any single lens component is not. 
The mass ratio that results from this approach is less strongly supportive of the major merger interpretation discussed below. Further exploration of the dynamical profile of the cluster for a more comprehensive 3D projection would help refine the measurement. In the following sections, we consider the measurement based on the total projected mass density enclosed in 80 kpc, for consistency with previous work.}

\begin{table*}
    \centering
    \begin{tabular}{cccccc}
    \hline
     &	Total  &	BCG subcluster  &	LRG subcluster &	 &	 	\\
        &	M($<$500 kpc) &	M$_{BCG}$ ($<80$ kpc)  &	M$_{LRG}$ ($<80$ kpc)&	Sub cluster ratio &	Galaxy fraction 	\\
Model   & [$10^{14}$ \msun]   &   [$10^{13}$ \msun]   &   [$10^{13}$ \msun]   &  M$_{BCG}$/M$_{LRG}$ & \\
  \hline
A &	$3.57^{+0.21}_{-0.25}$ &	$3.88^{+0.11}_{-0.11}$ &	$2.74^{+0.10}_{-0.08}$ &	$1.42^{+0.07}_{-0.07}$ &	$0.18^{+0.06}_{-0.05}$ 	\\
B &	$3.75^{+0.40}_{-0.34}$ &	$3.93^{+0.21}_{-0.14}$ &	$2.92^{+0.16}_{-0.23}$ &	$1.35^{+0.16}_{-0.08}$ &	$0.15^{+0.08}_{-0.03}$ 	\\
C &	$3.66^{+0.36}_{-0.36}$ &	$3.90^{+0.14}_{-0.14}$ &	$3.05^{+0.13}_{-0.27}$ &	$1.28^{+0.14}_{-0.04}$ &	$0.18^{+0.02}_{-0.08}$ 	\\
D &	$3.72^{+0.39}_{-0.26}$ &	$3.98^{+0.21}_{-0.06}$ &	$2.97^{+0.11}_{-0.21}$ &	$1.34^{+0.14}_{-0.03}$ &	$0.1\pm{0.01}$	\\
E &	$3.72^{+0.45}_{-0.18}$ &	$4.09^{+0.13}_{-0.21}$ &	$2.89^{+0.18}_{-0.17}$ &	$1.41^{+0.07}_{-0.10}$ &	$0.1\pm{0.01}$	\\
\hline
    \end{tabular}
    \caption{Total projected mass density (cylindrical mass) of the cluster as a whole, and the subclusters and their ratio, measured from the outputs of different lens models. Despite the different modeling choices, these measurements are in excellent agreement between models. Statistical uncertainties are 95\% confidence interval. The subcluster masses are measured within 80 kpc from the BCG and the brightest galaxy in the LRG group. The total cluster mass and galaxy fraction are measured within 500 kpc. Note that the galaxies component in Model D and E assumed a fixed scaling relation, resulting in a mass fraction of $\sim10$\% that was not constrained by the lens model. See \autoref{fig:masses} for a visualization, and \autoref{sec:disc} for a discussion of these results.}
    \label{table:masses}
\end{table*}

\begin{figure}
    \centering
    \includegraphics[width=3 in]{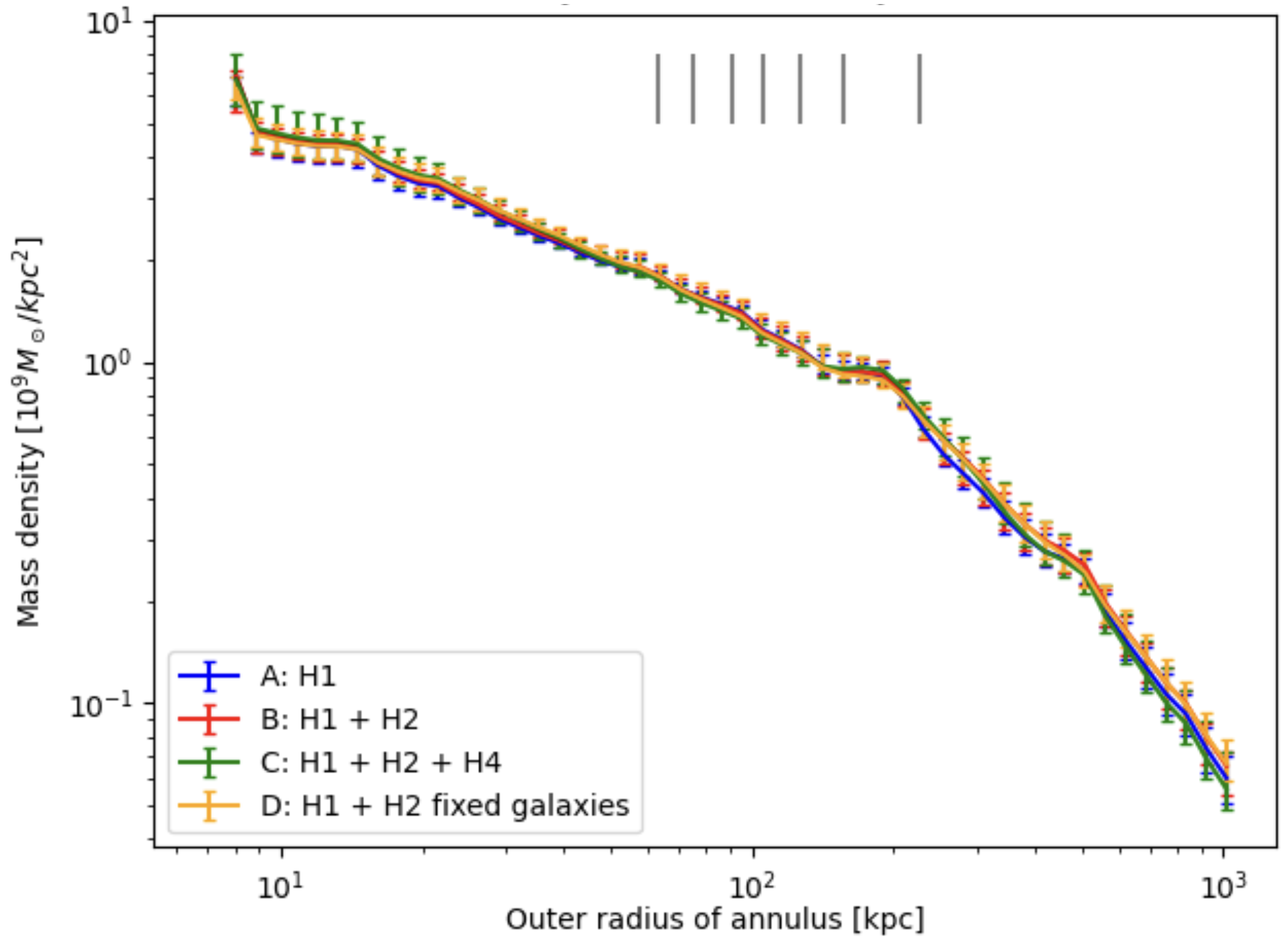}
    \caption{Total projected mass density profiles for the range of models investigated in this paper. The average mass densities are measured in annuli centered on the BCG. The ``bump'' at $\sim200$ kpc corresponds to the LRG structure. Uncertainties are 95\% confidence interval, derived from the MCMC sampling of the parameter space. \revision{The projected distances of the lensing constraints are indicated with gray ticks.}}
    \label{fig:mass-profile-total}
\end{figure}

\section{Discussion} \label{sec:disc}
\subsection{Evidence for a Major-merger Scenario}
\done{Our lensing analysis indicates that there are two major mass components in this cluster, which align with the distribution of cluster member galaxies. The two components have different galaxy content -- one is dominated by the BCG, and the other is made up of a group of a few smaller LRGs. This result agrees with the findings of \cite{Mahler2020}. }

\done{Also in agreement with \cite{Mahler2020}, we find that most of the relative motion between the two components is on the plane of the sky, based on new MUSE data that yielded spectroscopic redshifts of 32 cluster members. We found 21 cluster member galaxies within the MUSE footprint that are at $1.03<z<1.04$, including the BCG and nearby galaxies and seven of the galaxies in the LRG group. The median offset between the BCG and the LRG group is 135 \kms.  The MUSE spectroscopy that is presented in this paper will be used in a future dynamical investigation of this cluster.}

\cite{Dawson2012} presented four criteria for observing two subclusters during a dissociative major-merger event: 
(1) the two components have comparable mass, (2) the two components have a small impact parameter, (3) The cluster is observed during the period where cluster gas is significantly offset from the galaxies and dark matter, (4) The motion of the components is mostly transverse, so that the apparent angular separation of the gas from the galaxies and dark matter is maximized. 
In this paper, we demonstrate that \clustername\ satisfies criteria 1, 2, and 4. 
The mass ratio between the two major components of the cluster of $1:1.35$ (\autoref{sec:results}) implies that the two components have comparable mass, thus satisfying the first criterion. This conclusion is robust to modeling assumptions.  The separation on the plane of the sky is small, $\sim21''$, or $\sim170$ kpc, satisfying the second criterion. The similar redshift of the components satisfies the fourth. 
Further data are required to study the gas component and determine the dynamical state of the merger. This can be done with X-ray data: two distinct X-ray components associated with the two dark matter components would indicate that the cluster is in a premerger phase, and a single X-ray peak between the mass components would indicate a recent merger \citep[e.g.,][]{Clowe2006}. An analysis of the Chandra X-ray data obtained by program ID \#22800530 (PI: Mahler) will be presented in a future work. \revision{ Preliminary results indicate that the centroid of the X-ray emission is consistent with being between the two halos (\autoref{fig:mass-contours})}.

\subsection{Comparison to Previous Work}
\done{\cite{Mahler2020} published the first strong lensing analysis and lens model of \clustername, as described above, using ground-based imaging data and multislit spectroscopy and a single-band HST image. With the data available at the time, they were able to identify three sets of arcs to be used as lensing constraints, all of which appear between the two subclusters. 
The main improvements offered by the current work are the increased quantity and distribution of lensing constraints and the spectroscopic redshifts of numerous objects in the cluster core, enabled by the multiband HST imaging and VLT/MUSE IFU spectroscopy. 
The new data also facilitated a better selection of cluster member galaxies to be included in the lens model, increasing their number from 45 to 149. We note that the higher rms in the new model is expected because it has more arcs \citep{johnsonsharon}.}

\done{
To compare the masses of the two main subclusters between models, we used the model of \cite{Mahler2020} to compute the total enclosed mass within 80~kpc of the BCG and of the LRG group in the same way as was done in \autoref{sec:massresults} and estimated the 95\% confidence interval statistical uncertainties from the MCMC sampling. Obtaining these measurements directly from the lens model was necessary, since the reported numbers in \cite{Mahler2020} did not aggregate all the mass components, including the different cluster-scale halos and galaxy contribution in the BCG and group, and were reported with 68\% uncertainty. 
{The \cite{Mahler2020} fiducial model (model B in their Table~4, which has one cluster halo, DM1 free, optimized BCG)   yields 
\massBCGMB, in agreement with the new model, \massBCGB. For the LRG subcluster, the \cite{Mahler2020} fiducial model yields a higher mass than the new model, \massLRGsMB\ compared to \massLRGsB\ from our model B. The new model is in excellent agreement with models that allow more flexibility, e.g., model D in \cite{Mahler2020}, which allows for a second dark matter halo near the LRGs, and yields \massLRGsMD. }}

\done{
The mass within 500 kpc is in agreement with \cite{Mahler2020} within the error bars, although the new models predict a somewhat higher large-scale mass than the fiducial model of \cite{Mahler2020}, likely driven by the new evidence that the strong-lensing regime extends at least to the radius of arc 7 in the east. We reduced the statistical uncertainty on the total mass of the cluster (within 500 kpc) from {$\sim 16\%$}  in the fiducial model of \cite{Mahler2020} to {$\sim10\%$}. 
Owing to the new constraints in the east and west, the new model improved the statistical uncertainty on the positions of individual mass components. Most notably, the previous model, which had no constraints west of the group of LRGs, suffered from a high uncertainty in the position of H2, in particular, the R.A.\ could only be localized to within a $25\arcsec$ range. We localized the R.A.\ of H2 to within $\sim15\arcsec$. 
Additionally, all the models investigated in this paper that allow for a second cluster halo generally agree on the location of H2 within statistical uncertainties, indicating that the additional constraining power of the new system 4 (the westernmost arc, See \autoref{fig:constraints}) is driving this improvement, rather than modeling choices.}

\done{The main conclusion, that the system is composed of two subclusters of comparable mass with a ratio of $\sim 1: 1.35$, is in good agreement between the new analysis and previous work.}

\subsection{Mass composition and model degeneracies}\label{subsec:degeneracies}
\done{The primary output of gravitational lens modeling is the distribution of projected mass density in the lens plane. One of the strengths of parametric lens modeling algorithms such as \lenstool, which assume that the mass distribution is composed of the sum of several functional halos, is their ability to examine the relative contributions from different cluster components: the different major cluster halos and the galaxy-scale halos.  \autoref{fig:mass-profile} shows the contributions of the different mass components to the overall cluster mass and aids in comparing the different mass models to each other and pointing out model degeneracies.}

\done{Unsurprisingly, the peak in the galaxy component profile seen at $\sim200$ kpc coincides with the group of LRGs,  also visible in the total mass profiles as a bump at the same radius. As can be seen in the top line in \autoref{fig:mass-profile} and the profiles ploted in \autoref{fig:mass-profile-total}, the total projected mass profiles from the {four} models are nearly indistinguishable, indicating that modeling choices have little to no effect on the overall cluster radial profile. }

\done{As lensing is sensitive to the total mass regardless of its origin, lens models can distribute the mass between, e.g., the BCG and the main dark matter halos, or add or remove mass from cluster member galaxies by compensating elsewhere in the model. 
\autoref{fig:mass-profile} (deconstructed mass profiles) clearly shows this effect: a model with a lower galaxy component (D) compensates with a higher H2 mass. A model without H2 or H4 compensates by increasing the mass of H1.
This explains the observation that regardless of modeling choices (one, two, or three cluster halos, fixed or free galaxy scaling relations), the total projected mass density of the subclusters (\autoref{table:masses}, \autoref{fig:masses}) is consistent between models. }

\done{Similarly, the relative contribution from cluster member galaxies to the total mass of the cluster (measured within 500 kpc) is consistent between models that explored the parameter space of this property, predicting a mass fraction that ranges from $0.15^{+0.01}_{-0.02}$ (model B) to $0.18^{+0.02}_{-0.02}$ (model C). However, owing to degeneracies between model parameters, the lensing analysis cannot rule out a lower mass fraction -- models with a fixed galaxy contribution of $10$\% (models D/E) performed similarly well as models with free scaling relations parameters, indicating that this property is not well constrained by strong lensing alone in this cluster. }

\done{Examining the differences between the models in \autoref{fig:halos} (visualization of mass halos), \autoref{fig:mass-contours} (mass contours), and \autoref{table:best_fit_model} (model parameters) reveals more interesting insights about the effects of modeling choices and model flexibility on the overall mass distribution and the limitations of complex mass models in determining the relative contribution of mass components. }
\done{Including H4 in the lens model adds mass in the eastern region of the cluster core (\autoref{fig:halos}), which is required to satisfy the constraints of system 7. Models without H4 (models A, B and D),
add the needed mass in this region by slightly increasing the ellipticity and normalization of the major cluster component, H1 (e.g., compare the $e$ and $\sigma_0$ parameters of H1 of the different models in \autoref{table:best_fit_model}).  The effect of this compensation for the lowered flexibility in the eastern part of the core is seen in \autoref{fig:mass-contours}  as a deviation of the contours of model C from the other models, but only outside of the region where strong-lensing constraints exist, thus not significantly affecting or driving the performance of the models. The shape of the cluster mass distribution on a large scale cannot be constrained by strong lensing alone. Weak lensing can possibly help break the degeneracy between models B and C and help constrain the large-scale ellipticity of the mass distribution, since the effects of weak lensing are observed farther out from the cluster core. }

\done{The model degeneracies also come to light when comparing the new models to the models from \cite{Mahler2020}. That work evaluated models with H1 and a free BCG halo, with and without H2. Table 4 in that paper shows, for example, that the two models with similar rms have vastly different ratios of H1, H2, and BCG contributions -- the model with H1 fixed to the location of the BCG resulted with a very low BCG normalization, attributing its mass to the coincident cluster dark matter halo. As noted in \autoref{sec:halos}, the conclusion of our attempts to free the BCG parameters was that the added flexibility is not justified, as models with free BCG parameters do not perform better than models with the BCG using the same scaling relations as the other red-sequence galaxies. }

\done{These degeneracies indicate that we are unable to tightly constrain the relative contributions of mass components or the percentage of cluster mass contained in galaxies. We are also unable to constrain the shape of the model outside of where we have strong-lensing constraints, as the same results can be produced by adding mass or increasing ellipticity. Additional constraints provided by weak lensing could aid in this endeavor and offer more insight into the geometry of this cluster.}

\subsection{{Physical and statistical criteria of model selection}}
\done{We present in this analysis five lens models with different levels of complexity. To assess these models, we used two criteria (\autoref{sec:results}): the rms, which is directly linked to the goodness of fit, and the BIC, which is broadly used in Bayesian analyses to select between a finite set of models where overfitting is a concern (for previous work that used BIC in strong lens modeling analysis, see, e.g.,  \citealt{Mahler2018,Mahler2020,Acebron2022,Biviano2023}).
In this section, we discuss the use of the BIC criterion and contrast it with physical considerations for discriminating between models while keeping in mind that our lens modeling aims to reproduce the mass distribution of a cluster of galaxies. All models used the exact same number of constraints; i.e., $ln(n)=ln(40)$ is a constant in all models. Therefore, the BIC effectively quantifies the extent to which additional free parameters are able to maximize the $ln(L)$ \citep[for a definition of the likelihood function in \lenstool, we refer the reader to][]{Jullo2007, Niemiec2020}. It is to be noted that a difference in BIC of $>10$ is considered strong \citep{Kass1995,lorah2019} and a difference in BIC of $<2$ is not significant .}

\done{
Model B produced a lower rms than model A (rms = $0\farcs17$ vs.\ rms = $0\farcs20$); therefore it provides a better fit to the data that, while a small difference, could be argued to be significant. The lower BIC favors model A, which is the most simplistic of the five models under assessment, with one halo located near the BCG. Using the BIC straightforwardly (or naively) for selecting our model would mean that adding one more free cluster-scale potential (six free parameters) at the location of the LRGs is considered to be over-fitting, increasing the BIC from 32.9 (model A) to 50.9 (model B).  However, we can provide a physical argument that  model A has the largest offset between the center of the cluster-scale potential and the BCG as well as the largest position angle misalignment compared to every other model reported in this analysis. We can reason that the model is trying to compensate for the mass lacking in the LRG location with H1, to achieve the fixed total mass needed to satisfy the lensing constraints. As we demonstrated in \autoref{table:masses}, the projected mass within 80 kpc is consistent between models regardless of whether H2 is included. While it is expected that cluster halos and BCGs tend to align with each other \citep{Lambas1988,Ragone-Figueroa2020} and we are still well within the range of possibility even in model A, we stress that in every other model tested here, when given the flexibility to do so, H1 tends to better align with BCG morphology.}

\done{Comparing model B and model D offers insights into the ability of the BIC to consider the effect of allowing the scaling relation of cluster member galaxies to vary. The difference of about 7 in BIC estimates tends to slightly favor model D with a fixed scaling relation over model B, since it has a similar goodness of fit and two fewer parameters. We argue here that allowing the scaling relation to vary is essential for examining the degeneracy among the subcomponent of a model. Model B is better able to capture the underlying uncertainties in the distribution of dark matter  within the cluster itself. We again stress that the strong lensing configuration is giving a robust estimate of the total amount of mass within 80 kpc regardless of the model. However, the model with the galaxy scaling relation free to vary is probing a more diverse parameter space. Fixing these parameters could lead to an incomplete account of the statistical error budget.}

\done{Model C presents the best fit to the data with the lowest rms of $0\farcs14$; it is, however, disfavored by having the highest BIC. Naively, this model could be discarded on the basis of its high BIC (73.7), suggesting that it overfits the data by including halo H4. However, an inspection of the projected mass density contours of the cluster as presented in \autoref{fig:mass-contours} reveals that the change in shape is most prominent at the outskirts. As previously mentioned, the contours in the outskirts of all other models (A, B, and D, which do not include H4) are showing very elliptical mass distribution, matching the ones from H1 at the very edge, beyond the strong-lensing region. Model C has a more complex mass distribution with a significantly lower ellipticity on large scales. As previously mentioned we cannot with strong lensing data alone constrain the morphology of this cluster far beyond the strong lensing regime but we again stress that the sole use of BIC is preventing exploration of the model and should be considered alongside physical considerations when selecting favored models.}

\section{Summary} \label{sec:conc}

We present new multiband HST imaging and a strong-lensing analysis of \clusterfullname, a galaxy cluster at redshift $z=\clusterz$. We build upon and improve the initial lens model constructed by \cite{Mahler2020} by identifying additional lensing constraints, including two new lensed galaxies with spectroscopic redshifts, for a total of five high-confidence lensed sources with multiple images, and several candidate arcs. 
We used the resulting lens models to estimate the mass ratio of the two main components of \clustername, with the goal of confirming its classification as a major merger. 

We also report in this paper the serendipitous discovery of a background group of three doubled-peaked \lya\ emitting galaxies at $z=\zsystemX$, magnified by a factor of 2--4 by the cluster. \lya\ emission is also detected in three of the strongly-lensed galaxies, at $z=\zsystemIII$, $z=\zsystemIV$, and $z=\zsystemVII$.

\done{The multiband HST imaging and MUSE spectroscopy resulted in the robust identification of constraints in previously underconstrained areas of the lens, most importantly, east of the BCG and west of the LRGs, 
which allows us to better constrain the projected mass density distribution of the cluster.
The galaxy distribution and the lensing geometry indicate that the cluster has at least two major cores. 
We found that a mass model with two cluster-scale halos adequately reproduces the lensing evidence while keeping the number of free parameters low. 
We explored a possibility that the lens plane is more complex by allowing a flexible third halo (H4). A model with three cluster-scale halos does produce a lower image plane rms, however, a BIC analysis determined that the slight improvement in rms is not justified by the number of added parameters. We therefore selected {model B} as a sufficient representation of the lens plane given the available constraints. Further constraints on the large scale ellipticity of the cluster from weak lensing can help break the degeneracy between models.}

{We used the derived lens models to calculate the total projected mass density enclosed within a radius of 80~kpc centered on each of the subclusters, the ``BCG'' core and the ``LRG'' core. 
The subcluster masses are \massBCGB\ and \massLRGsB\ for the BCG core and LRG core, respectively. The total mass of the cluster, measured as projected (cylindrical) mass within 500~kpc from the BCG, is \massclusterB. 
We find that 
despite making different assumptions on the number of subhalos, all models have consistent mass ratio results, yielding a mass ratio between 
{1:1.28 and 1:1.42}, and support the classification of \clustername\ as a major merger, in agreement with \cite{Mahler2020}. Combined with the small separation of the sky ($\sim 170$ kpc) and the small radial velocity offset ($135$ \kms), we conclude that the cluster is observed during a major merger on the plane of the sky.

Future analysis using X-ray data and advanced dynamical study of the cluster core could classify this cluster as a dissociative merger, which would make it a unique target to probe large-scale structure formation, cluster assembly, and galaxy evolution in the densest nodes of the cosmic web.

\begin{acknowledgments}
\done{This research is based on observations made with the NASA/ESA Hubble Space Telescope obtained from the Space Telescope Science Institute, which is operated by the Association of Universities for Research in Astronomy, Inc., under NASA contract NAS 5–26555. These observations are associated with programs GO-16425, GO-13412. Support for program \# 16425 was provided by NASA through a grant from the Space Telescope Science Institute, which is operated by the Association of Universities for Research in Astronomy, Inc., under NASA contract NAS 5-03127.
Based on observations collected at the European Southern Observatory under ESO programme 106.21JF.001 obtained from the ESO Science Archive Facility.
The South Pole Telescope program is supported by the National Science Foundation (NSF) through awards OPP-1852617 and 2332483. Partial support is also provided by the Kavli Institute of Cosmological Physics at the University of Chicago. Work at Argonne National Lab is supported by UChicago Argonne LLC, Operator of Argonne National Laboratory (Argonne). Argonne, a U.S. Department of Energy Office of Science Laboratory, is operated under contract no. DE-AC02-06CH11357.}
\end{acknowledgments}

\done{
\facilities{HST(ACS, WFC3), VLT(MUSE)}
\software{Source Extractor \citep{Bertin1996}; \lenstool\ \citep{Jullo2007};  AstroDrizzle \citep{astrodrizzle}; Matlab \citep{MATLAB}; Astropy \citep{astropy} ; MATLAB Astronomy and Astrophysics Toolbox \citep[MAAT;][]{Ofek2014}}}

\newpage

\bibliography{sample631.bib}{}

\begin{thebibliography}{}
\expandafter\ifx\csname natexlab\endcsname\relax\def\natexlab#1{#1}\fi
\providecommand{\url}[1]{\href{#1}{#1}}
\providecommand{\dodoi}[1]{doi:~\href{http://doi.org/#1}{\nolinkurl{#1}}}
\providecommand{\doeprint}[1]{\href{http://ascl.net/#1}{\nolinkurl{http://ascl.net/#1}}}
\providecommand{\doarXiv}[1]{\href{https://arxiv.org/abs/#1}{\nolinkurl{https://arxiv.org/abs/#1}}}

\bibitem[{{Acebron} {et~al.}(2022){Acebron}, {Grillo}, {Bergamini}, {Caminha}, {Tozzi}, {Mercurio}, {Rosati}, {Brammer}, {Meneghetti}, {Nonino}, \& {Vanzella}}]{Acebron2022}
{Acebron}, A., {Grillo}, C., {Bergamini}, P., {et~al.} 2022, \aap, 668, A142, \dodoi{10.1051/0004-6361/202244836}

\bibitem[{{Astropy Collaboration} {et~al.}(2018){Astropy Collaboration}, {Price-Whelan}, {Sip{\H{o}}cz}, {G{\"u}nther}, {Lim}, {Crawford}, {Conseil}, {Shupe}, {Craig}, {Dencheva}, {Ginsburg}, {VanderPlas}, {Bradley}, {P{\'e}rez-Su{\'a}rez}, {de Val-Borro}, {Aldcroft}, {Cruz}, {Robitaille}, {Tollerud}, {Ardelean}, {Babej}, {Bach}, {Bachetti}, {Bakanov}, {Bamford}, {Barentsen}, {Barmby}, {Baumbach}, {Berry}, {Biscani}, {Boquien}, {Bostroem}, {Bouma}, {Brammer}, {Bray}, {Breytenbach}, {Buddelmeijer}, {Burke}, {Calderone}, {Cano Rodr{\'\i}guez}, {Cara}, {Cardoso}, {Cheedella}, {Copin}, {Corrales}, {Crichton}, {D'Avella}, {Deil}, {Depagne}, {Dietrich}, {Donath}, {Droettboom}, {Earl}, {Erben}, {Fabbro}, {Ferreira}, {Finethy}, {Fox}, {Garrison}, {Gibbons}, {Goldstein}, {Gommers}, {Greco}, {Greenfield}, {Groener}, {Grollier}, {Hagen}, {Hirst}, {Homeier}, {Horton}, {Hosseinzadeh}, {Hu}, {Hunkeler}, {Ivezi{\'c}}, {Jain}, {Jenness}, {Kanarek}, {Kendrew}, {Kern}, {Kerzendorf}, {Khvalko}, {King}, {Kirkby}, {Kulkarni},
  {Kumar}, {Lee}, {Lenz}, {Littlefair}, {Ma}, {Macleod}, {Mastropietro}, {McCully}, {Montagnac}, {Morris}, {Mueller}, {Mumford}, {Muna}, {Murphy}, {Nelson}, {Nguyen}, {Ninan}, {N{\"o}the}, {Ogaz}, {Oh}, {Parejko}, {Parley}, {Pascual}, {Patil}, {Patil}, {Plunkett}, {Prochaska}, {Rastogi}, {Reddy Janga}, {Sabater}, {Sakurikar}, {Seifert}, {Sherbert}, {Sherwood-Taylor}, {Shih}, {Sick}, {Silbiger}, {Singanamalla}, {Singer}, {Sladen}, {Sooley}, {Sornarajah}, {Streicher}, {Teuben}, {Thomas}, {Tremblay}, {Turner}, {Terr{\'o}n}, {van Kerkwijk}, {de la Vega}, {Watkins}, {Weaver}, {Whitmore}, {Woillez}, {Zabalza}, \& {Astropy Contributors}}]{astropy}
{Astropy Collaboration}, {Price-Whelan}, A.~M., {Sip{\H{o}}cz}, B.~M., {et~al.} 2018, \aj, 156, 123, \dodoi{10.3847/1538-3881/aabc4f}

\bibitem[{{Bacon} {et~al.}(2016){Bacon}, {Piqueras}, {Conseil}, {Richard}, \& {Shepherd}}]{Bacon2016}
{Bacon}, R., {Piqueras}, L., {Conseil}, S., {Richard}, J., \& {Shepherd}, M. 2016, {MPDAF: MUSE Python Data Analysis Framework}, Astrophysics Source Code Library, record ascl:1611.003

\bibitem[{{Bacon} {et~al.}(2010){Bacon}, {Accardo}, {Adjali}, {Anwand}, {Bauer}, {Biswas}, {Blaizot}, {Boudon}, {Brau-Nogue}, {Brinchmann}, {Caillier}, {Capoani}, {Carollo}, {Contini}, {Couderc}, {Daguis{\'e}}, {Deiries}, {Delabre}, {Dreizler}, {Dubois}, {Dupieux}, {Dupuy}, {Emsellem}, {Fechner}, {Fleischmann}, {Fran{\c{c}}ois}, {Gallou}, {Gharsa}, {Glindemann}, {Gojak}, {Guiderdoni}, {Hansali}, {Hahn}, {Jarno}, {Kelz}, {Koehler}, {Kosmalski}, {Laurent}, {Le Floch}, {Lilly}, {Lizon}, {Loupias}, {Manescau}, {Monstein}, {Nicklas}, {Olaya}, {Pares}, {Pasquini}, {P{\'e}contal-Rousset}, {Pell{\'o}}, {Petit}, {Popow}, {Reiss}, {Remillieux}, {Renault}, {Roth}, {Rupprecht}, {Serre}, {Schaye}, {Soucail}, {Steinmetz}, {Streicher}, {Stuik}, {Valentin}, {Vernet}, {Weilbacher}, {Wisotzki}, \& {Yerle}}]{Bacon2010}
{Bacon}, R., {Accardo}, M., {Adjali}, L., {et~al.} 2010, in Society of Photo-Optical Instrumentation Engineers (SPIE) Conference Series, Vol. 7735, Ground-based and Airborne Instrumentation for Astronomy III, ed. I.~S. {McLean}, S.~K. {Ramsay}, \& H.~{Takami}, 773508, \dodoi{10.1117/12.856027}

\bibitem[{{Bacon} {et~al.}(2017){Bacon}, {Conseil}, {Mary}, {Brinchmann}, {Shepherd}, {Akhlaghi}, {Weilbacher}, {Piqueras}, {Wisotzki}, {Lagattuta}, {Epinat}, {Guerou}, {Inami}, {Cantalupo}, {Courbot}, {Contini}, {Richard}, {Maseda}, {Bouwens}, {Bouch{\'e}}, {Kollatschny}, {Schaye}, {Marino}, {Pello}, {Herenz}, {Guiderdoni}, \& {Carollo}}]{Bacon2017}
{Bacon}, R., {Conseil}, S., {Mary}, D., {et~al.} 2017, \aap, 608, A1, \dodoi{10.1051/0004-6361/201730833}

\bibitem[{{Bertin} \& {Arnouts}(1996)}]{Bertin1996}
{Bertin}, E., \& {Arnouts}, S. 1996, \aaps, 117, 393, \dodoi{10.1051/aas:1996164}

\bibitem[{{Biviano} {et~al.}(2023){Biviano}, {Pizzuti}, {Mercurio}, {Sartoris}, {Rosati}, {Ettori}, {Girardi}, {Grillo}, {Caminha}, \& {Nonino}}]{Biviano2023}
{Biviano}, A., {Pizzuti}, L., {Mercurio}, A., {et~al.} 2023, \apj, 958, 148, \dodoi{10.3847/1538-4357/acf832}

\bibitem[{{Bleem} {et~al.}(2015){Bleem}, {Stalder}, {de Haan}, {Aird}, {Allen}, {Applegate}, {Ashby}, {Bautz}, {Bayliss}, {Benson}, {Bocquet}, {Brodwin}, {Carlstrom}, {Chang}, {Chiu}, {Cho}, {Clocchiatti}, {Crawford}, {Crites}, {Desai}, {Dietrich}, {Dobbs}, {Foley}, {Forman}, {George}, {Gladders}, {Gonzalez}, {Halverson}, {Hennig}, {Hoekstra}, {Holder}, {Holzapfel}, {Hrubes}, {Jones}, {Keisler}, {Knox}, {Lee}, {Leitch}, {Liu}, {Lueker}, {Luong-Van}, {Mantz}, {Marrone}, {McDonald}, {McMahon}, {Meyer}, {Mocanu}, {Mohr}, {Murray}, {Padin}, {Pryke}, {Reichardt}, {Rest}, {Ruel}, {Ruhl}, {Saliwanchik}, {Saro}, {Sayre}, {Schaffer}, {Schrabback}, {Shirokoff}, {Song}, {Spieler}, {Stanford}, {Staniszewski}, {Stark}, {Story}, {Stubbs}, {Vanderlinde}, {Vieira}, {Vikhlinin}, {Williamson}, {Zahn}, \& {Zenteno}}]{Bleem2015}
{Bleem}, L.~E., {Stalder}, B., {de Haan}, T., {et~al.} 2015, \apjs, 216, 27, \dodoi{10.1088/0067-0049/216/2/27}

\bibitem[{{Bleem} {et~al.}(2020){Bleem}, {Bocquet}, {Stalder}, {Gladders}, {Ade}, {Allen}, {Anderson}, {Annis}, {Ashby}, {Austermann}, {Avila}, {Avva}, {Bayliss}, {Beall}, {Bechtol}, {Bender}, {Benson}, {Bertin}, {Bianchini}, {Blake}, {Brodwin}, {Brooks}, {Buckley-Geer}, {Burke}, {Carlstrom}, {Rosell}, {Carrasco Kind}, {Carretero}, {Chang}, {Chiang}, {Citron}, {Moran}, {Costanzi}, {Crawford}, {Crites}, {da Costa}, {de Haan}, {De Vicente}, {Desai}, {Diehl}, {Dietrich}, {Dobbs}, {Eifler}, {Everett}, {Flaugher}, {Floyd}, {Frieman}, {Gallicchio}, {Garc{\'\i}a-Bellido}, {George}, {Gerdes}, {Gilbert}, {Gruen}, {Gruendl}, {Gschwend}, {Gupta}, {Gutierrez}, {Halverson}, {Harrington}, {Henning}, {Heymans}, {Holder}, {Hollowood}, {Holzapfel}, {Honscheid}, {Hrubes}, {Huang}, {Hubmayr}, {Irwin}, {James}, {Jeltema}, {Joudaki}, {Khullar}, {Klein}, {Knox}, {Kuropatkin}, {Lee}, {Li}, {Lidman}, {Lowitz}, {MacCrann}, {Mahler}, {Maia}, {Marshall}, {McDonald}, {McMahon}, {Melchior}, {Menanteau}, {Meyer}, {Miquel}, {Mocanu},
  {Mohr}, {Montgomery}, {Nadolski}, {Natoli}, {Nibarger}, {Noble}, {Novosad}, {Padin}, {Palmese}, {Parkinson}, {Patil}, {Paz-Chinch{\'o}n}, {Plazas}, {Pryke}, {Ramachandra}, {Reichardt}, {Remolina Gonz{\'a}lez}, {Romer}, {Roodman}, {Ruhl}, {Rykoff}, {Saliwanchik}, {Sanchez}, {Saro}, {Sayre}, {Schaffer}, {Schrabback}, {Serrano}, {Sharon}, {Sievers}, {Smecher}, {Smith}, {Soares-Santos}, {Stark}, {Story}, {Suchyta}, {Tarle}, {Tucker}, {Vanderlinde}, {Veach}, {Vieira}, {Wang}, {Weller}, {Whitehorn}, {Wu}, {Yefremenko}, \& {Zhang}}]{Bleem2020}
{Bleem}, L.~E., {Bocquet}, S., {Stalder}, B., {et~al.} 2020, \apjs, 247, 25, \dodoi{10.3847/1538-4365/ab6993}

\bibitem[{{Bocquet} {et~al.}(2019){Bocquet}, {Dietrich}, {Schrabback}, {Bleem}, {Klein}, {Allen}, {Applegate}, {Ashby}, {Bautz}, {Bayliss}, {Benson}, {Brodwin}, {Bulbul}, {Canning}, {Capasso}, {Carlstrom}, {Chang}, {Chiu}, {Cho}, {Clocchiatti}, {Crawford}, {Crites}, {de Haan}, {Desai}, {Dobbs}, {Foley}, {Forman}, {Garmire}, {George}, {Gladders}, {Gonzalez}, {Grandis}, {Gupta}, {Halverson}, {Hlavacek-Larrondo}, {Hoekstra}, {Holder}, {Holzapfel}, {Hou}, {Hrubes}, {Huang}, {Jones}, {Khullar}, {Knox}, {Kraft}, {Lee}, {von der Linden}, {Luong-Van}, {Mantz}, {Marrone}, {McDonald}, {McMahon}, {Meyer}, {Mocanu}, {Mohr}, {Morris}, {Padin}, {Patil}, {Pryke}, {Rapetti}, {Reichardt}, {Rest}, {Ruhl}, {Saliwanchik}, {Saro}, {Sayre}, {Schaffer}, {Shirokoff}, {Stalder}, {Stanford}, {Staniszewski}, {Stark}, {Story}, {Strazzullo}, {Stubbs}, {Vanderlinde}, {Vieira}, {Vikhlinin}, {Williamson}, \& {Zenteno}}]{Bocquet2019}
{Bocquet}, S., {Dietrich}, J.~P., {Schrabback}, T., {et~al.} 2019, \apj, 878, 55, \dodoi{10.3847/1538-4357/ab1f10}

\bibitem[{{Brada{\v{c}}} {et~al.}(2008){Brada{\v{c}}}, {Allen}, {Treu}, {Ebeling}, {Massey}, {Morris}, {von der Linden}, \& {Applegate}}]{Bradac2008}
{Brada{\v{c}}}, M., {Allen}, S.~W., {Treu}, T., {et~al.} 2008, \apj, 687, 959, \dodoi{10.1086/591246}

\bibitem[{{Brada{\v{c}}} {et~al.}(2006){Brada{\v{c}}}, {Clowe}, {Gonzalez}, {Marshall}, {Forman}, {Jones}, {Markevitch}, {Randall}, {Schrabback}, \& {Zaritsky}}]{Bradac2006}
{Brada{\v{c}}}, M., {Clowe}, D., {Gonzalez}, A.~H., {et~al.} 2006, \apj, 652, 937, \dodoi{10.1086/508601}

\bibitem[{{Caminha} {et~al.}(2023){Caminha}, {Grillo}, {Rosati}, {Liu}, {Acebron}, {Bergamini}, {Caputi}, {Mercurio}, {Tozzi}, {Vanzella}, {Demarco}, {Frye}, {Rosani}, \& {Sharon}}]{Caminha2023}
{Caminha}, G.~B., {Grillo}, C., {Rosati}, P., {et~al.} 2023, \aap, 678, A3, \dodoi{10.1051/0004-6361/202244897}

\bibitem[{{Clowe} {et~al.}(2006){Clowe}, {Brada{\v{c}}}, {Gonzalez}, {Markevitch}, {Randall}, {Jones}, \& {Zaritsky}}]{Clowe2006}
{Clowe}, D., {Brada{\v{c}}}, M., {Gonzalez}, A.~H., {et~al.} 2006, \apjl, 648, L109, \dodoi{10.1086/508162}

\bibitem[{{Dawson} {et~al.}(2012){Dawson}, {Wittman}, {Jee}, {Gee}, {Hughes}, {Tyson}, {Schmidt}, {Thorman}, {Brada{\v{c}}}, {Miyazaki}, {Lemaux}, {Utsumi}, \& {Margoniner}}]{Dawson2012}
{Dawson}, W.~A., {Wittman}, D., {Jee}, M.~J., {et~al.} 2012, \apjl, 747, L42, \dodoi{10.1088/2041-8205/747/2/L42}

\bibitem[{{El{\'\i}asd{\'o}ttir} {et~al.}(2007){El{\'\i}asd{\'o}ttir}, {Limousin}, {Richard}, {Hjorth}, {Kneib}, {Natarajan}, {Pedersen}, {Jullo}, \& {Paraficz}}]{Eliastottir07}
{El{\'\i}asd{\'o}ttir}, {\'A}., {Limousin}, M., {Richard}, J., {et~al.} 2007, arXiv e-prints, arXiv:0710.5636, \dodoi{10.48550/arXiv.0710.5636}

\bibitem[{{Fakhouri} \& {Ma}(2008)}]{Fakhouri2008}
{Fakhouri}, O., \& {Ma}, C.-P. 2008, \mnras, 386, 577, \dodoi{10.1111/j.1365-2966.2008.13075.x}

\bibitem[{{Fruchter} \& {et al.}(2010)}]{astrodrizzle}
{Fruchter}, A.~S., \& {et al.} 2010, in 2010 Space Telescope Science Institute Calibration Workshop, 382--387

\bibitem[{{Fumagalli} {et~al.}(2016){Fumagalli}, {Cantalupo}, {Dekel}, {Morris}, {O'Meara}, {Prochaska}, \& {Theuns}}]{Fumagalli2016}
{Fumagalli}, M., {Cantalupo}, S., {Dekel}, A., {et~al.} 2016, \mnras, 462, 1978, \dodoi{10.1093/mnras/stw1782}

\bibitem[{{Fumagalli} {et~al.}(2017){Fumagalli}, {Mackenzie}, {Trayford}, {Theuns}, {Cantalupo}, {Christensen}, {Fynbo}, {M{\o}ller}, {O'Meara}, {Prochaska}, {Rafelski}, \& {Shanks}}]{Fumagalli2017}
{Fumagalli}, M., {Mackenzie}, R., {Trayford}, J., {et~al.} 2017, \mnras, 471, 3686, \dodoi{10.1093/mnras/stx1896}

\bibitem[{{Gladders} \& {Yee}(2000)}]{gladdersyee2000}
{Gladders}, M.~D., \& {Yee}, H.~K.~C. 2000, \aj, 120, 2148, \dodoi{10.1086/301557}

\bibitem[{{Harvey} {et~al.}(2015){Harvey}, {Massey}, {Kitching}, {Taylor}, \& {Tittley}}]{Harvey2015}
{Harvey}, D., {Massey}, R., {Kitching}, T., {Taylor}, A., \& {Tittley}, E. 2015, Science, 347, 1462, \dodoi{10.1126/science.1261381}

\bibitem[{Inc.(2022)}]{MATLAB}
Inc., T.~M. 2022, MATLAB version: 9.13.0 (R2022b),  Natick, Massachusetts, United States: The MathWorks Inc.
\newblock \url{https://www.mathworks.com}

\bibitem[{{Johnson} \& {Sharon}(2016)}]{johnsonsharon}
{Johnson}, T.~L., \& {Sharon}, K. 2016, \apj, 832, 82, \dodoi{10.3847/0004-637X/832/1/82}

\bibitem[{{Jullo} {et~al.}(2007){Jullo}, {Kneib}, {Limousin}, {El{\'\i}asd{\'o}ttir}, {Marshall}, \& {Verdugo}}]{Jullo2007}
{Jullo}, E., {Kneib}, J.~P., {Limousin}, M., {et~al.} 2007, New Journal of Physics, 9, 447, \dodoi{10.1088/1367-2630/9/12/447}

\bibitem[{Kass \& Raftery(1995)}]{Kass1995}
Kass, R.~E., \& Raftery, A.~E. 1995, Journal of the American Statistical Association, 90, 773, \dodoi{10.1080/01621459.1995.10476572}

\bibitem[{{Lambas} {et~al.}(1988){Lambas}, {Groth}, \& {Peebles}}]{Lambas1988}
{Lambas}, D.~G., {Groth}, E.~J., \& {Peebles}, P.~J.~E. 1988, \aj, 95, 996, \dodoi{10.1086/114695}

\bibitem[{{Lorah} \& {Womack}(2019)}]{lorah2019}
{Lorah}, J., \& {Womack}, A. 2019, Behavior Research Methods, 51, 440, \dodoi{10.3758/s13428-018-1188-3}

\bibitem[{{Mahler} {et~al.}(2018){Mahler}, {Richard}, {Cl{\'e}ment}, {Lagattuta}, {Schmidt}, {Patr{\'\i}cio}, {Soucail}, {Bacon}, {Pello}, {Bouwens}, {Maseda}, {Martinez}, {Carollo}, {Inami}, {Leclercq}, \& {Wisotzki}}]{Mahler2018}
{Mahler}, G., {Richard}, J., {Cl{\'e}ment}, B., {et~al.} 2018, \mnras, 473, 663, \dodoi{10.1093/mnras/stx1971}

\bibitem[{{Mahler} {et~al.}(2020){Mahler}, {Sharon}, {Gladders}, {Bleem}, {Bayliss}, {Calzadilla}, {Floyd}, {Khullar}, {McDonald}, {Remolina Gonz{\'a}lez}, {Schrabback}, {Stark}, \& {van den Busch}}]{Mahler2020}
{Mahler}, G., {Sharon}, K., {Gladders}, M.~D., {et~al.} 2020, \apj, 894, 150, \dodoi{10.3847/1538-4357/ab886b}

\bibitem[{{Markevitch} {et~al.}(2004){Markevitch}, {Gonzalez}, {Clowe}, {Vikhlinin}, {Forman}, {Jones}, {Murray}, \& {Tucker}}]{Markevitch2004}
{Markevitch}, M., {Gonzalez}, A.~H., {Clowe}, D., {et~al.} 2004, \apj, 606, 819, \dodoi{10.1086/383178}

\bibitem[{{Menanteau} {et~al.}(2012){Menanteau}, {Hughes}, {Sif{\'o}n}, {Hilton}, {Gonz{\'a}lez}, {Infante}, {Barrientos}, {Baker}, {Bond}, {Das}, {Devlin}, {Dunkley}, {Hajian}, {Hincks}, {Kosowsky}, {Marsden}, {Marriage}, {Moodley}, {Niemack}, {Nolta}, {Page}, {Reese}, {Sehgal}, {Sievers}, {Spergel}, {Staggs}, \& {Wollack}}]{Menanteau2012}
{Menanteau}, F., {Hughes}, J.~P., {Sif{\'o}n}, C., {et~al.} 2012, \apj, 748, 7, \dodoi{10.1088/0004-637X/748/1/7}

\bibitem[{{Niemiec} {et~al.}(2020){Niemiec}, {Jauzac}, {Jullo}, {Limousin}, {Sharon}, {Kneib}, {Natarajan}, \& {Richard}}]{Niemiec2020}
{Niemiec}, A., {Jauzac}, M., {Jullo}, E., {et~al.} 2020, \mnras, 493, 3331, \dodoi{10.1093/mnras/staa473}

\bibitem[{{Ofek}(2014)}]{Ofek2014}
{Ofek}, E.~O. 2014, {MAAT: MATLAB Astronomy and Astrophysics Toolbox}, Astrophysics Source Code Library, record ascl:1407.005.
\newblock \doeprint{1407.005}

\bibitem[{{Ragone-Figueroa} {et~al.}(2020){Ragone-Figueroa}, {Granato}, {Borgani}, {De Propris}, {Garc{\'\i}a Lambas}, {Murante}, {Rasia}, \& {West}}]{Ragone-Figueroa2020}
{Ragone-Figueroa}, C., {Granato}, G.~L., {Borgani}, S., {et~al.} 2020, \mnras, 495, 2436, \dodoi{10.1093/mnras/staa1389}

\bibitem[{{Randall} {et~al.}(2008){Randall}, {Markevitch}, {Clowe}, {Gonzalez}, \& {Brada{\v{c}}}}]{Randall2008}
{Randall}, S.~W., {Markevitch}, M., {Clowe}, D., {Gonzalez}, A.~H., \& {Brada{\v{c}}}, M. 2008, \apj, 679, 1173, \dodoi{10.1086/587859}

\bibitem[{Schwarz(1978)}]{Schwarz1978}
Schwarz, G. 1978, The Annals of Statistics, 6, 461 , \dodoi{10.1214/aos/1176344136}

\bibitem[{{Soto} {et~al.}(2016){Soto}, {Lilly}, {Bacon}, {Richard}, \& {Conseil}}]{Soto2016}
{Soto}, K.~T., {Lilly}, S.~J., {Bacon}, R., {Richard}, J., \& {Conseil}, S. 2016, \mnras, 458, 3210, \dodoi{10.1093/mnras/stw474}

\bibitem[{{Weilbacher}(2015)}]{Weilbacher2015}
{Weilbacher}, P. 2015, in Science Operations 2015: Science Data Management, 1, \dodoi{10.5281/zenodo.34658}

\bibitem[{{Wittman} {et~al.}(2018){Wittman}, {Golovich}, \& {Dawson}}]{Wittman2018}
{Wittman}, D., {Golovich}, N., \& {Dawson}, W.~A. 2018, \apj, 869, 104, \dodoi{10.3847/1538-4357/aaee77}

\end{thebibliography}
\bibliographystyle{aasjournal}

\end{document}